\documentclass[prl,twocolumn,showpacs,a4paper]{revtex4-1}
\usepackage{epsfig}
\usepackage{amsmath}
\usepackage{graphicx}
\usepackage[usenames,dvipsnames]{color}
\usepackage{hyperref}
\usepackage{bbm}
\usepackage{umoline}

\definecolor{myblue}{rgb}{0,0,0.75}

\begin{document}

\title{Implementation of the Dicke lattice model in hybrid quantum system arrays}

\author{L. J. Zou$^1$, D. Marcos$^2$, S. Diehl$^3$, S. Putz$^4$,  J. Schmiedmayer$^4$, J. Majer$^4$, P. Rabl$^4$  }
\affiliation{$^1$Department of Physics, Harvard University, Cambridge, Massachusetts 02138, USA}
\affiliation{$^2$Institute for Quantum Optics and Quantum Information of the 
Austrian Academy of Sciences, A-6020 Innsbruck, Austria}
\affiliation{$^3$Institute for Theoretical Physics, University of Innsbruck, A-6020 Innsbruck, Austria}
\affiliation{$^4$Vienna Center for Quantum Science and Technology, Atominstitut, Vienna University of Technology,
Stadionallee 2, 1020 Vienna, Austria}

\date{\today}

\begin{abstract}
Generalized Dicke models can be implemented in hybrid quantum systems built from ensembles of nitrogen-vacancy (NV) centers in diamond coupled to superconducting microwave cavities. By engineering cavity assisted Raman transitions between two spin states
of the NV defect, a fully tunable model for collective light-matter interactions in the ultra-strong coupling limit can be obtained.
Our analysis of the resulting non-equilibrium phases for a single cavity and for coupled cavity arrays shows that different superradiant phase transitions can be observed using existing experimental technologies, even in the presence of large inhomogeneous broadening of the spin ensemble. The phase diagram of the Dicke lattice model displays distinct features induced by dissipation, which can serve as a genuine experimental signature for phase transitions in driven open quantum systems.
\end{abstract}

\maketitle 

The Dicke model (DM)~\cite{Dicke} was first introduced to describe the collective coupling of $N$ two-level atoms to a single optical mode, and has  developed into a prototype model for collective quantum phenomena in atomic and solid-state systems~\cite{BrandesPR2005}. Most prominently, 
the DM predicts a \emph{superradiant} phase transition (SRT) \cite{HeppLieb, WangHioe, Emary} when the collective atom-field coupling reaches the \emph{ultra-strong coupling} regime and becomes comparable to the optical and atomic frequencies. While the existence of this transition for atoms coupled directly to an optical mode is subject of ongoing debates~\cite{Rzazewski, Knight, Keeling,DeLiberato,VukicsPRL2014}, effective DMs can be implemented, for example, using tailored Raman couplings in driven cold-atom systems~\cite{DimerPRA2007,NagyPRL2010,BhaseenPRA2012,RitschRMP2013,IvanovJPhysB2013,Lesanovsky}. In these systems, the non-equilibrium SRT~\cite{DimerPRA2007,OztopNJP2012}  has recently  been observed~\cite{Esslinger,HamnerNatComm2014,Baden2014}, which represents an important step towards more detailed investigations of phase transitions in open quantum systems. Yet a true many-body generalization of these models with multiple \emph{independent} atomic and optical degrees of freedom 
\cite{Strack2011,Buchhold2013,Schiro2012,Hartmann2008,Tomadin2010,CarusottoRMP2013} still faces considerable experimental and theoretical challenges.

\begin{figure}[b]
 \begin{center}
 \includegraphics[width=0.45\textwidth]{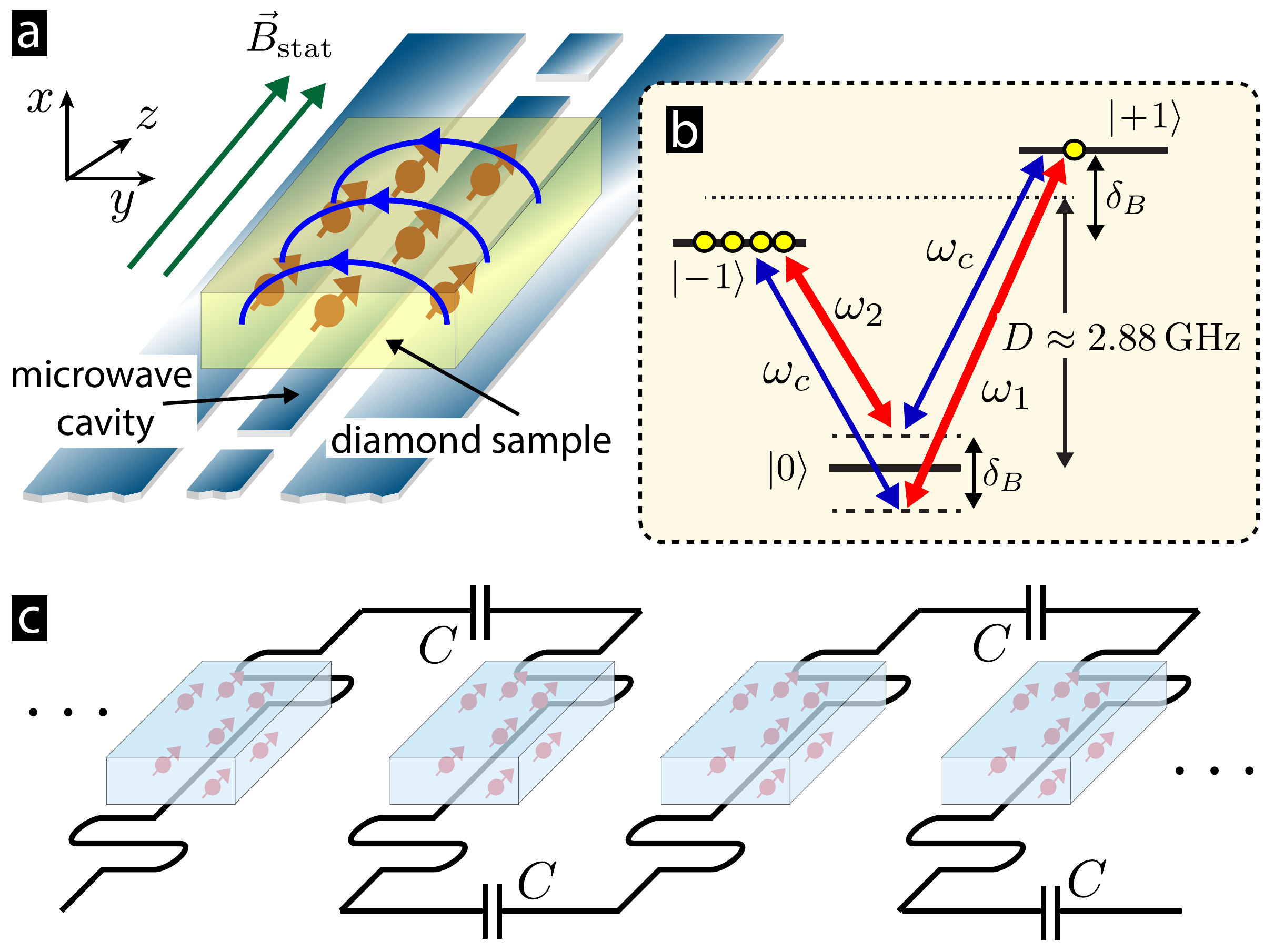} 
 \caption{(color online). a) A planar microwave cavity is coupled to an ensemble of NV center spins in a diamond sample placed on top. b) Level diagram of the NV center ground state. The blue arrows indicate the coupling to the quantized cavity field of frequency $\omega_c$. Two additional classical microwave fields with frequencies $\omega_{1,2}$ are used to implement Raman transitions between the two excited spin states $|\pm1 \rangle$. c) An array of capacitively-coupled cavity-spin ensemble systems is used for the implementation of the Dicke lattice model. }  
 \label{fig:Setup}
 \end{center}
 \end{figure}

In this work we describe a new approach for realizing generalized DMs, by using atomic~\cite{VerduPRL2009}, molecular~\cite{RablPRL2006} or solid-state spin ensembles~\cite{ImamogluPRL2009,WesenbergPRL2009,Marcos10} coupled to superconducting microwave cavities [see Fig.~\ref{fig:Setup} a)]. Such \emph{hybrid quantum systems}~\cite{Bensky2011,XiangRMP2013} have originally been proposed for quantum information processing applications, and strong collective interactions between microwave photons and solid-state spin ensembles have already been observed~\cite{SchusterPRL2010,KuboPRL2010,AmsussPRL2011,Semba,RanjanPRL2013,ProbstPRL2013,Tabuchi2014}. 
Compared to optical or all superconducting circuit realizations~\cite{ChenPRA2007,Ciuti,HouckNatPhys2012},  the current approach allows us to combine large ensembles of (almost) identical spins with high-quality microwave resonators that can be easily coupled together to form large arrays~\cite{UnderwoodPRA2012}. For the example of nitrogen-vacancy (NV) spin ensembles in diamond, we describe the implementation of effective light-matter interactions in the ultra-strong coupling regime and evaluate the resulting non-equilibrium phases for a single spin ensemble and for coupled  spin ensembles-cavity arrays. Our findings demonstrate that hybrid quantum systems provide a realistic platform for implementing Dicke-type lattice models and for studying characteristic phenomena of non-equilibrium quantum systems in various 1D or 2D configurations.

\emph{Model.} 
We consider a setup as shown in Fig.~\ref{fig:Setup} a), where the quantized magnetic field of a superconducting microwave cavity is coupled to an ensemble of NV center spins in diamond. 
The NV defect~\cite{NVReview} has a spin $S=1$ triplet ground state with a zero-field splitting of $D\approx 2.88 $ GHz between the $|m_s=0\rangle$ and the $|m_s=\pm1\rangle$ spin states [see Fig.~\ref{fig:Setup} b)].
The spin Hamiltonian for a single center is $H_{\rm NV} =  \hbar  D S_z^2 +\mu_B g_s \mathcal{B}_z S_z+H_\Omega(t)$, where $g_s\simeq 2$, $\mu_B$ is the Bohr magneton and $\mathcal{B}_z=\vec n_z \cdot  \vec B_{\rm stat}$ is the component of a static bias field, $ \vec B_{\rm stat}$, along the NV symmetry axis, $\vec n_z$. We assume that $ \vec B_{\rm stat}$ is homogeneous over the extent of the sample and oriented such, that all NV centers experience the same Zeeman splitting $\delta_B=2 \mu_B g_s |\vec B_{\rm stat}|/(\sqrt{3}\hbar)$ between the $|\pm 1\rangle$ states~\cite{SuppInfo}. Finally, $H_\Omega(t)=\sum_{\xi=\pm1,n=1,2} \frac{\Omega_n}{2} \left(Ê  e^{i\omega_n t} |0\rangle\langle \xi|Ê  +   {\rm H.c.}\right)$ accounts for spin rotations, which are driven by two classical microwave fields of frequencies $\omega_{1,2}\sim D$ [see Fig.~\ref{fig:Setup} b)] and Rabi frequencies $\Omega_1$ and $\Omega_2$, respectively.

The cavity is modeled as a single-mode harmonic oscillator with frequency $\omega_c\sim D$ and annihilation operator $a$. A NV center located at $\vec r_i$, 
couples to the quantized cavity field, $\vec B_c(\vec r)=\vec B_0(\vec r) (a+a^\dag)$, with a vacuum Rabi frequency $g_0^i= g_s \mu_B |\vec B_{0,i}^\perp|/(\sqrt{3}\hbar) $, where $\vec B_0(\vec r)$ is the magnetic field associated with a single photon and $\vec B_{0,i}^\perp\sim \vec B_0(\vec r_i)$ is the relevant field component orthogonal to the NV symmetry axis (see \cite{SuppInfo} for details).
Putting everything together, the Hamiltonian for a single cavity coupled to a spin ensemble is $(\hbar=1)$ 
\begin{equation}\label{eq:Hfull}
\begin{split}Ê
H=  &\omega_c a^\dag a + \sum_{i, \xi=\pm 1} \left(D+ \xi \frac{(\delta_B+\delta_i)}{2}\right) |\xi\rangle_i\langle \xi|  \\
&+\sum_{i, \xi=\pm1}   g_0^i\left( a+ a^\dag\right) \left( |0\rangle_i\langle \xi|Ê  +  |\xi\rangle_i\langle 0|\right) + H_{\Omega}^i(t),
\end{split} 
\end{equation}
where the random offsets $\delta_i\sim $ MHz account for the inhomogeneous broadening of the spin ensemble. Inhomogeneous shifts arise from local strain, couplings to other impurity spins and hyperfine interactions~\cite{SuppInfo}, and can be assumed to be static over the relevant timescales.

For typical values, $g_0\sim 10$ Hz, the single spin coupling as well as the collective coupling $G_0\simeq g_0\sqrt{\mathcal{N}}$ for an ensemble of $\mathcal{N}\sim 10^{12}$ spins are much smaller than  $\omega_c\sim D$.  To achieve ultra-strong coupling conditions, we engineer an effective model~\cite{DimerPRA2007}, where the two excited spin states $|\pm1\rangle$ are coupled via two-photon Raman transitions that involve the cavity and one of the classical fields. For the choice  $\omega_c\approx D$ and $\omega_{1,2}\approx D\pm \delta_B$ indicated in Fig.~\ref{fig:Setup} b), the two possible transitions from $|-1\rangle$ to $|+1\rangle$ either involve the absorption or the emission of a cavity photon and result in both Jaynes-Cummings and anti-Jaynes-Cummings interactions.  For   $|\delta_B|\gg G_0, |\Omega_n|,|\delta_i|$, and with all  NV centers initially prepared in state $|-1\rangle$, we can eliminate the state $|0\rangle$ and obtain an effective Hamiltonian~\cite{SuppInfo}
 \begin{equation}\label{eq:Heff}
\begin{split}Ê
H_{\rm eff}=Ê& \Delta_c a^\dag a +  \sum_i  \left[  \frac{\Delta_s^i}{2}Ê+ \lambda_i  a^\dag a \right] (\sigma_z^i+1) \\
&+ \sum_i  \left(g_1^i  a  +  g_2^i a^\dag \right)\sigma_-^i + {\rm H.c.}, 
\end{split} 
\end{equation}
where the $\sigma_{z,\pm}^i$ are Pauli operators acting on the states $|\pm1\rangle_i$,  $g_n^i = g_0^i\Omega_n^i/\delta_B$ and $\lambda_i=2(g_0^i)^2/\delta_B$. The effective cavity and spin frequencies, $\Delta_c = \omega_c - (\omega_1+\omega_2)/2-\sum_i \lambda_i$, and $\Delta^i_s = \delta_B-(\omega_1-\omega_2)/2 + \delta_i - (|\Omega_1^i|^2+|\Omega_2^i|^2)/(3\delta_B)$, are 
two-photon detunings, which can be adjusted by an appropriate choice of $\omega_{1,2}$. 
Finally, we include a finite photon loss rate $2\kappa$ and model the dissipative system dynamics by a master equation 
\begin{equation}
\dot \rho = -i [H_{\rm eff} ,\rho]Ê + \kappa \left( 2a \rho a^\dag -a^\dag a \rho - \rho a^\dag a \right). 
\end{equation}
Note that this approach is valid in our current setting, since strong coupling is only achieved within the effective model while interactions between actual photonic and spin excitations are still weak~\cite{RidolfoPRL2012}. 
At cryogenic temperatures the spin $T_1$ time is several seconds~\cite{AmsussPRL2011,Harrison2006} and spin decay can be neglected. 

\emph{The inhomogeneous Dicke model. ---} For a homogenous system where $\Delta_s^i=\Delta_s$ and $g_1^i=g_2^i=g$, Eq.~\eqref{eq:Heff} reproduces the standard DM with an additional Stark shift term $\sim \lambda$. At zero temperature and $\kappa=0$, this model exhibits a second-order quantum phase transition from a normal phase with $\langle a\rangle=0$ to a superradiant phase with $\langle a\rangle \neq0$ at a critical coupling $G_{\rm crit} =\sqrt{\Delta_c \Delta_s/4}$, where $G=g\sqrt{\mathcal{N}}$ \cite{BrandesPR2005}. For $\lambda>0$ the Stark shift term slightly modifies the superradiant phase, but does not change the transition point~\cite{BhaseenPRA2012,SuppInfo}.
In the present setting we encounter the opposite scenario, where the $g_0^i\sim B_0(\vec r_i)$ vary strongly over the diamond sample, and the inhomogeneous broadening of the spins, $\gamma_s$, can even exceed the other frequency scales, $\gamma_s > G,\Delta_s, \Delta_c$. This means that the $\Delta_s^i$ can be close to zero or even negative and the existence of a stable normal phase and a sharp SRT is a priori not  evident. 
To analyze the DM under these conditions,  we extend the approach of Ref.~\cite{GotoPRA2008} and divide spins into subgroups of $N_\mu$ spins with approximately the same parameters $\Delta_s^i\simeq \Delta_\mu$, $g_i\simeq g_\mu$ and $\lambda_i\simeq \lambda_\mu$. Then
\begin{equation}\label{eq:HDM_mulit}
\begin{split}
H_{\rm eff} = &\Delta_c a^\dag a + \sum_\mu \left( \Delta_\mu +2\lambda_\mu  a^\dag a \right) \left(J_\mu^z+\frac{N_\mu}{2}\right) \\
&+ \sum_\mu \frac{G_\mu}{\sqrt{N_\mu}}(a+a^\dag)(J_\mu^-+J_\mu^+),
\end{split}
\end{equation} 
where $J_\mu^{z}=1/2\sum_{i\in \mu} \sigma_z^i $ and $J_\mu^{\pm} =\sum_{i\in \mu} \sigma_\pm^i $ are collective spin operators and $G_\mu =\sqrt{N_\mu}g_\mu$ is the collective coupling for each subgroup. Assuming a relatively large total number of spins, the individual subgroups may still be treated as collective spins with $J_\mu=N_\mu/2\gg1$.

In the limit $G_\mu\rightarrow 0$, the stationary expectation values are $\langle a\rangle=\langle J_\mu^-\rangle = 0$ and $\langle J_\mu^z\rangle=-N_\mu/2$ and the system is in the normal phase. We use a Holstein-Primakoff approximation~\cite{DimerPRA2007,SuppInfo} to represent spin excitations on top of this fully polarized state by bosonic operators, i.e. $J_z^\mu \simeq b_\mu^\dag b_\mu - N_\mu/2$ and  $J_\mu^-\simeq \sqrt{N_\mu} b_\mu$.  
From the resulting quadratic Hamiltonian we derive a set of coupled equations for the mean amplitudes,
 \begin{eqnarray}
 \langle \dot a\rangle   &=& -(i\Delta_c+\kappa)  \langle a\rangle - i \sum_\mu  G_\mu \left( \langle b_\mu\rangle   +  \langle b_\mu^\dag\rangle \right),\\
\langle \dot b_\mu\rangle   &=& -i\Delta_\mu \langle b_\mu\rangle    - i G_\mu  \left( \langle a\rangle  + \langle a^\dag\rangle \right),
 \end{eqnarray}
which can be written in a matrix from as $ \dot{\vec{v}} = {\bf M} \vec v$, where 
$\vec v=(\langle a\rangle ,\langle a^\dag\rangle ,\langle b_1\rangle ,\langle b_1^\dag\rangle , \langle b_2\rangle ,\langle b_2^\dag\rangle, \dots)^T$. 
In the normal phase all eigenvalues of ${\bf M}$ have a negative real part and all system excitations are damped. The SRT occurs when the real part of one of the eigenvalues changes sign and the normal phase becomes unstable. This occurs when~\cite{SuppInfo}
 \begin{equation}\label{eq:PTCondition1}
 \lim_{\epsilon\rightarrow 0}  \sum_\mu  \frac{4 G^2_\mu \Delta_c \Delta_\mu }{(\Delta_c^2+\kappa^2)(\Delta_\mu^2+\epsilon^2)}=1.
 \end{equation}
Assuming a sufficiently dense distribution of collective spin modes, we  introduce a normalized spectral density $\rho(\omega)=G^{-2}\sum_\mu G_\mu^2 \delta(\omega-\Delta_\mu)$~\cite{WesenbergPRA2011,SandnerPRA2012}, where  $G=\sqrt{\sum_\mu G_\mu^2}$. Then Eq.~\eqref{eq:PTCondition1} can be written as
 \begin{equation}\label{eq:GcritGeneral}
\frac{4G^2}{\Delta_c\bar\Delta_s(1+\kappa^2/\Delta_c^2)}  \times \mathcal{P} \int d\omega\, \frac{\bar \Delta_s}{\omega} \rho(\omega)=1,
 \end{equation}
 where $\mathcal{P}$ denotes the Cauchy principal value and $\bar \Delta_s=\langle \Delta_\mu\rangle$ is the average spin frequency.  Eq.~\eqref{eq:GcritGeneral} generalizes the Dicke phase transition point for systems with photon loss~\cite{DimerPRA2007} and arbitrary coupling~\cite{GotoPRA2008} and frequency distributions.

  \begin{figure}
 \begin{center}
 \includegraphics[width=0.48\textwidth]{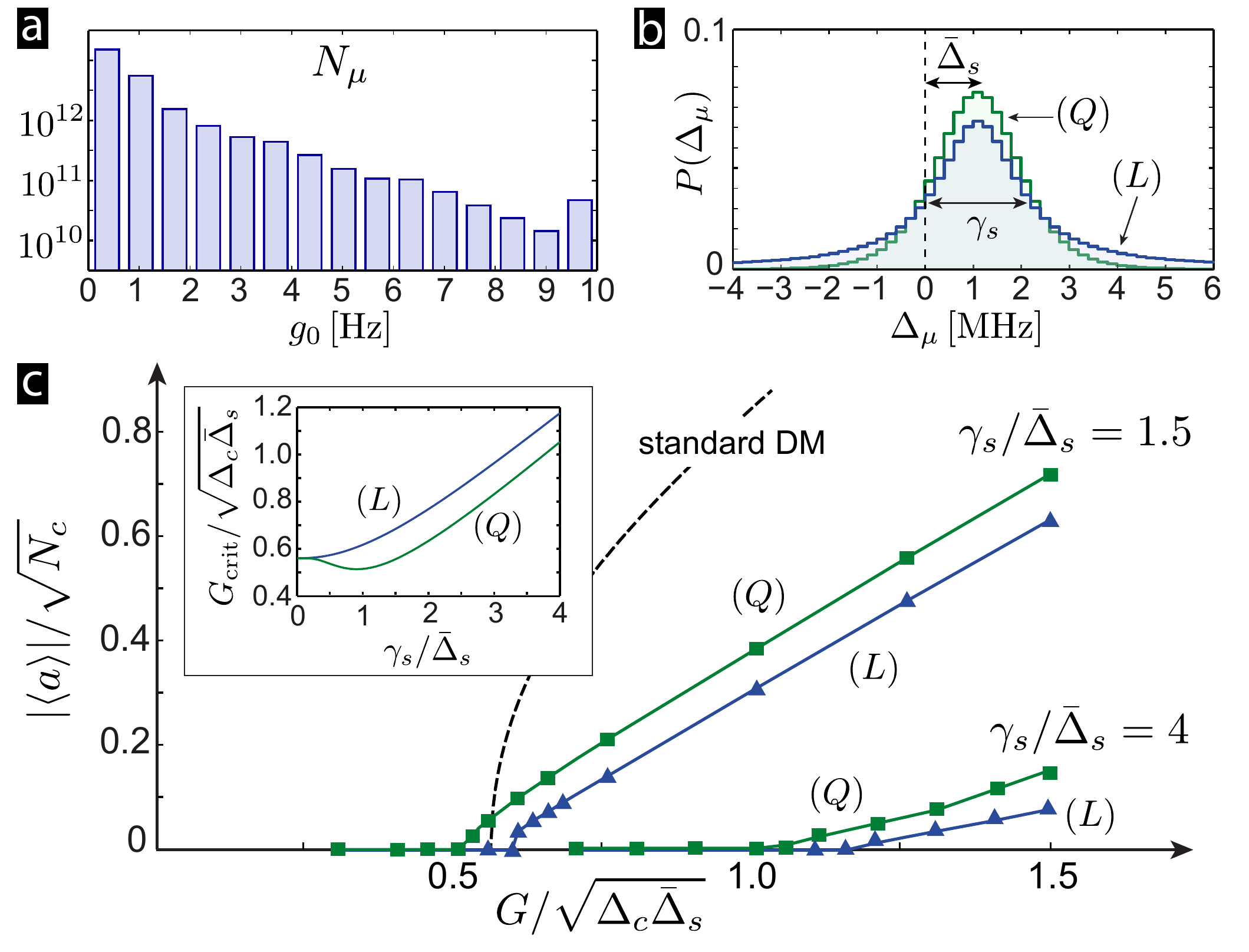} 
 \caption{(color online). Superradiant phase transition of the inhomogeneous DM. a) Total number of spins $N_\mu$ with a bare spin-cavity coupling $g_0$ for a diamond sample specified in~\cite{SuppInfo}. All spins with higher couplings are included in the last bin. b) Discretized distribution of spin frequencies $\Delta_\mu$. The two curves approximate a Lorentzian and $q$-Gaussian ($q=1.3$) distribution with  the same width at half maximum (FWHM) $\gamma_s$ and centered around $\bar \Delta_s$. c) The steady state field expectation value  $|\langle a\rangle|$ is plotted as a function of the collective coupling $G$ and for different values of $\gamma_s$. The dashed line indicates the result for the standard homogeneous DM, including the cavity decay, but without the Stark shift. The inset shows the critical coupling strength for the two different distributions calculated from Eq.~\eqref{eq:PTCondition1}. The parameters used in c) are derived from the coupling and frequency distributions shown in a) and b), $\Omega/\delta_B\leq 0.2$ and $\Delta_c=\bar \Delta_s=2\kappa=1$ MHz. For these parameters $N_c\approx1.2 \times 10^{13}$ and $\sum_\mu \lambda_\mu N_\mu \approx 1.2$ MHz. See text and~\cite{SuppInfo} for more details.
  }  
 \label{fig:PhaseTransition}
 \end{center}
 \end{figure}

\emph{Discussion.---} Fig.~\ref{fig:PhaseTransition} c) shows the steady state value of $|\langle a\rangle|/\sqrt{N_c}$, where $N_c=(\sum_\mu g_\mu N_\mu)^2/G^2$ is the characteristic photon number, as a function of $G$ and for different frequency distributions $P(\Delta_\mu=\bar\Delta_s+\delta_\mu)$. 
For this plot we have numerically integrated the semi-classical equations of motion for the mean values of $\langle a\rangle(t)$ and $\langle J_\mu^{z,\pm}\rangle(t)$ and assumed homogeneous classical fields, $\Omega_{n}^i=\Omega$.
In this case  $g_\mu\sim g_0^\mu\Omega /\delta_B$ and $\rho(\omega)\equiv P(\omega)$. 
From the distribution of bare couplings 
$g_{0}^\mu$ evaluated for a typical electrode configuration [cf. Fig.~\ref{fig:PhaseTransition} a)], $\delta_B=100$ MHz and $0 < \Omega < 20$ MHz,   
we obtain a maximal collective Raman coupling $G=G_0\Omega/\delta_B\approx 1.5$ MHz, which is consistent with experimentally observed values of $G_0\approx 10$ MHz~\cite{KuboPRL2010,AmsussPRL2011}. All parameters used for this calculation are detailed in~\cite{SuppInfo}.

For a Lorentzian distribution, $P(\omega)=(\gamma_s/2\pi)/(\omega^2+\gamma_s^2/4)$, Eq.~\eqref{eq:GcritGeneral} predicts a critical coupling strength
\begin{equation}
G_{\rm crit} = \sqrt{\frac{\Delta_c\bar \Delta_s}{4}\left(1+\frac{\kappa^2}{\Delta_c^2}\right)\left(1+\frac{\gamma_s^2}{4\bar \Delta_s^2}\right) },
\end{equation} 
which shows that the SRT occurs even in the regime of large frequency broadening, $\gamma_s >\bar \Delta_s$. Indeed, by optimizing $\Delta_c$ and $\bar \Delta_s$, the minimal requirement for observing the SRT is a strong collective cooperativity,
$\mathcal{C}_{\mathcal{N}} =\frac{2G^2}{\kappa\gamma_s} >1$.  In current experiments, where $\gamma_s \approx 20$ MHz~\cite{SandnerPRA2012,Putz2014}, this condition can be achieved for the above mentioned couplings and $\kappa=0.1$ MHz. Further,  in those experiments $P(\omega)$ resembles a $q-$Gaussian distribution, which is shown as a second example in Fig.~\ref{fig:PhaseTransition} b). This slightly narrower distribution leads to lower values of $G_{\rm crit}$, which can even  lie below the critical coupling obtained for a homogeneous sample.

{\it The Dicke lattice model. ---} Compared to optical cavities, microwave resonators can be fabricated with almost identical frequencies  and coupled together capacitively to form large 1D or 2D arrays~\cite{UnderwoodPRA2012}. Therefore, when combined with spin ensembles as described above, the current setting provides a feasible approach to implement lattice-generalizations of the DM as illustrated in Fig.~\ref{fig:Setup} c). For $N_L$ coupled cavities the resulting Dicke lattice model (DLM) is described by the Hamiltonian
\begin{equation} \label{eq:DickeLattice}
\begin{split}
H_{\rm DLM} &=\Delta_c \sum_{\ell=1}^{N_L} a^\dagger_{\ell} a_{\ell}   -t\sum_{\ell=1}^{N_L-1}(a^\dagger_\ell a_{\ell+1}+a_\ell a^\dag_{\ell+1})\\
&+ \sum_{\ell=1}^{N_L} \Delta_s  J^z_\ell + \sum_{\ell=1}^{N_L} \frac{G}{\sqrt{\mathcal{N}}}(J^+_{\ell}+J^-_\ell)(a_\ell+a_\ell^\dag),
\end{split}
\end{equation}
where $t$ is the coupling strength between neighboring cavities. For simplicity we have in Eq.~\eqref{eq:DickeLattice} represented each spin ensemble by a single collective spin $\vec J_\ell$ and neglected the Stark shift term $\sim\lambda$, which does not significantly change the relevant properties of this model. 

 \begin{figure}
 \begin{center}
 \includegraphics[width=0.48\textwidth]{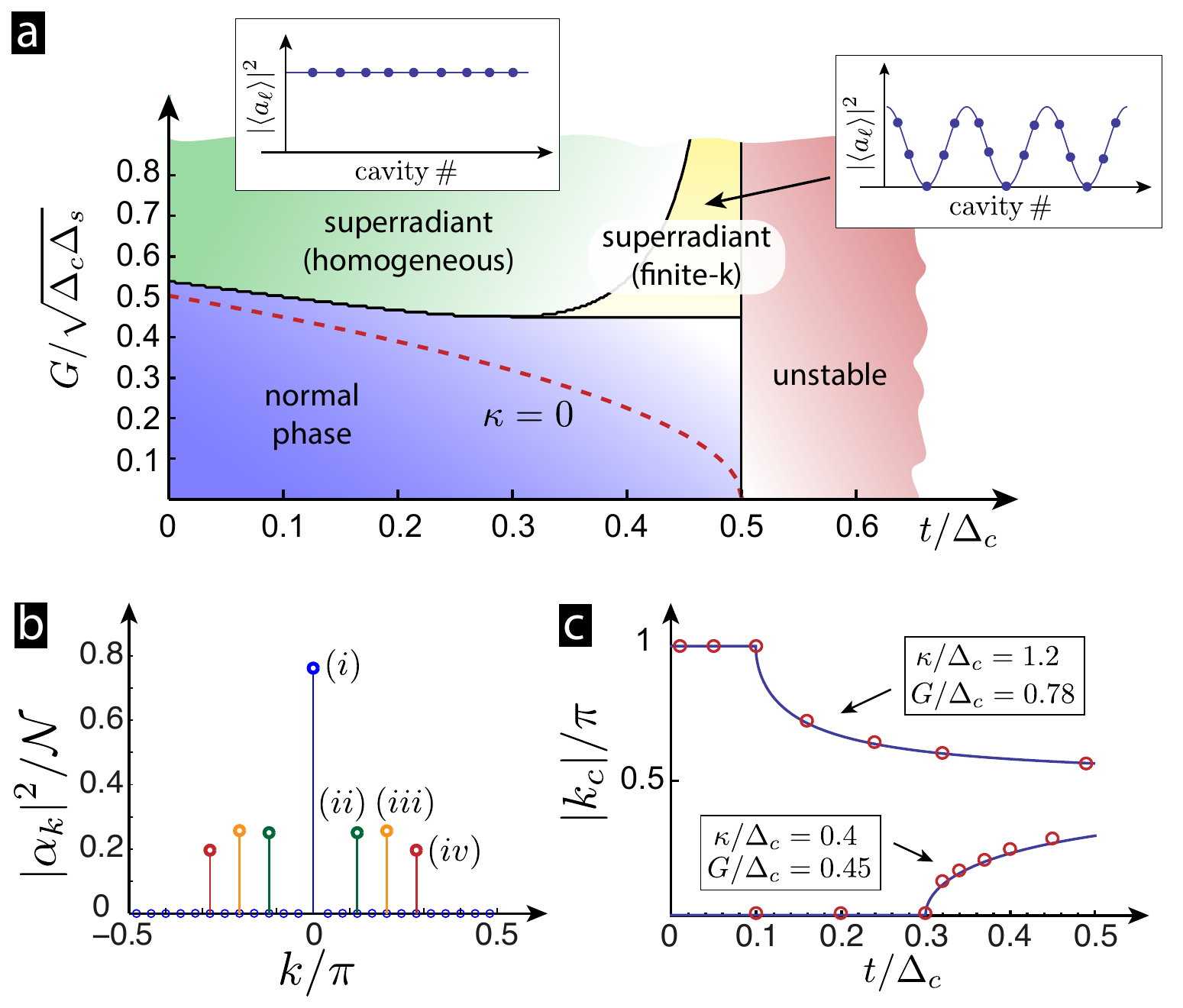} 
 \caption{(color online). a) Steady state phase diagram of the DLM for $\kappa/\Delta_c=0.4$ and $\Delta_c=\Delta_s>0$. The insets indicate the mean cavity photon numbers in the homogeneous and `finite-k'  superradiant phases. b) The value of $|\alpha_k|^2$, where $\alpha_k=1/\sqrt{N_L}\sum_{\ell=1}^{N_L} e^{ik \ell} \alpha_\ell$, is plotted for a lattice of $N_L=50$ site with periodic boundary conditions and $G=0.45\Delta_c>G_{\rm crit}$. For each curve, where $t/\Delta_c=0.3\, (i), 0.32\, (ii), 0.37\, (iii),0.45\, (iv)$, only one or two values at $k=\pm k_c$ are significantly different from zero.  c) The value of $k_c$ is plotted as a function of $t$ for $\kappa<\Delta_c$ and $\kappa>\Delta_c$. The solid lines represent the analytic result discussed in the text and the red dots are the numerically evaluated values. }  
 \label{fig:PhaseDiagram}
 \end{center}
 \end{figure}
 
As in the case of a single cavity, $H_{\rm DLM}$ represents an effective model for the underlying driven interaction described by Eq.~\eqref{eq:Hfull} and Fig.~\ref{fig:PhaseDiagram} shows the resulting non-equilibrium phase diagram of the DLM for $\Delta_c=\Delta_s>0$ and a finite photon loss $2\kappa<\Delta_c$. The different phases are characterized by the stationary values of $\langle a_\ell\rangle$, which we obtain from numerically solving the coupled semi-classical equations for $\langle a_\ell\rangle$, $\langle J_\ell^-\rangle$ and $\langle J_\ell^z\rangle$ combined with a fluctuation analysis~\cite{SuppInfo}. For $t\rightarrow 0$ the cavities are almost decoupled and as we increase $G$ we recover the standard SRT from the normal phase to a \emph{homogeneous} superradiant phase with $\langle a_\ell\rangle=\alpha \neq 0$.  For larger $t$ the coupled cavities form a frequency band $\Delta_k=\Delta_c-2t\cos(k)$, with quasi momentum  $k\in (-\pi,\pi]$.  
This reduces the frequency of the lowest $k=0$ photonic mode, which then leads to an instability at a reduced critical coupling  
 \begin{equation} 
G_{\rm crit}=\frac{1}{2}\sqrt{\Delta_s(\Delta_c -2t)\left[1+\frac{\kappa^2}{(\Delta_c-2t)^2}\right]}.
\end{equation} 
For the range of tunneling parameters $|\Delta_c-\kappa|< 2t < \Delta_c$, a new transition appears at a fixed critical coupling $G_{\rm crit}=\sqrt{\kappa\Delta_s/2}$.  This transition is driven by fluctuations with a finite quasi momentum $k_c=\arccos(t_c/t)$, where $t_c=(\Delta_c-\kappa)/2$, and results in a superradiant phase with a spatially varying field expectation value $\langle a_\ell\rangle \simeq \alpha \cos(\phi_0+k_c\ell )$, $\alpha \in \mathbbm{C}$. In a homogeneous lattice the random offset $\phi_0$ obtained in each experimental run breaks translation invariance. For larger losses, $\kappa>\Delta_c+2t$, the system always favors anti-ferromagnetic ordering, i.e.,  $k_c=\pi$. Note that the transition to a \emph{finite-$k$} superradiant phase is absent in the equilibrium phase diagram and may be seen as a genuine non-equilibrium feature of our model~\cite{CrossHohenbergRevModPhys}. It is related to the fact that in a dissipative system the occurrence of an unstable mode does not necessarily coincide with one of the system excitation frequencies going to zero. Similar effects of pattern formation in driven open quantum systems have been identified in~\cite{DiehlPRL2010,LeePRL2013}

Finally, for $2t>\Delta_c$, one or more of the photonic frequencies $\Delta_k$ are negative and the normal phase becomes unstable for arbitrarily  small values of $G$. 
The origin of this instability can be understood as follows: For $G\ll |\Delta_c|,|\Delta_s|$ the coupling term $a J^+$ exchanges photonic and spin excitations with an energy penalty of $\Delta_s-\Delta_k$ and combined with the photon decay, this process stabilizes the normal phase. In contrast, the coupling term $a^\dag J^+$ simultaneously creates one photonic and one spin excitation with total energy $\Delta_k+\Delta_s$. Since only the photon decays, this process overall populates the spin mode. From simple energy arguments we see that, whenever $\Delta_k$ and $\Delta_s$ have opposite signs, this second process is more favorable and destabilizes the normal phase for arbitrary small $G$. 
Our numerical results~\cite{SuppInfo} confirm this intuitive picture, and in this unstable regime we observe very small values of $\alpha_\ell$, while at the same time the expectation values of  $\langle J_\ell^-\rangle$ and $\langle J_\ell^z\rangle$ exhibit large amplitude oscillations with no significant damping on the timescales of interest.      

 \emph{Conclusions and outlook.---}ÊIn summary we have shown that hybrid quantum system arrays offer a  realistic platform for studying the Dicke model, and in particular its lattice generalizations,  in a natural way. The spatial continuum of degrees of freedom generated here for arrays in one or two dimensions paves the way for the experimental exploration of non-equilibrium phases and phase transitions in driven open systems. One key feature of the driven open lattice Dicke model identified here is the existence of a superradiant phase with additional spontaneous translation symmetry breaking, which does not have an immediate counterpart in equilibrium.  Our approach can be adapted to other atomic and solid state systems with multiple spin components~\cite{VerduPRL2009,RablPRL2006,ProbstPRL2013} and spin ensembles coupled to nonlinear superconducting circuits~\cite{HuemmerPRA2012,Qiu2014}.

\emph{Acknowledgments.} The authors thank P. Bertet, P. Bushev, D. Krimer, S. Rotter, C. L. Yu and P. Zoller for stimulating discussions and input on this project. This work was supported by the European Project SIQS and the Austrian Science Fund (FWF) through SFB FOQUS, the Doctoral School Solids4Fun (Project W1243), Project No. F4006-N16 and the START grants Y 591-N16 (PR) and Y 581-N16 (SD). J.M. acknowledges support by the TOP grant of the TU Wien. L. J. Z. acknowledges financial support from the Tsinghua Xuetang Talents Program 
and thanks IQOQI for hospitality.

\widetext

\newpage\newpage
\begin{center}
{\bf SUPPLEMENTARY MATERIAL}
\end{center}
\vspace{0.5cm}

This supplemental material contains additional details on the model and the derivation of various results stated in the main part of the paper.  In Sec.~{\bf I} 
we first present a detailed discussion of the magnetic coupling between a microwave cavity and a NV spin ensemble and in Sec.~{\bf II} 
we then derive the effective Dicke model for a single cavity.  In Sec.~{\bf III} 
we analyze the superradiant phase transition of the inhomogeneous Dicke model with arbitrary coupling and frequency distributions. A detailed discussion of the parameters which are used for the numerical simulations in the main text is given in Sec.~{\bf IV} 
Finally, in Sec.~{\bf V} 
we derive the stationary phases of the Dicke lattice model.

%
%

\section{{\bf I.} An ensemble of NV centers coupled to a transmission line cavity}\label{sec:Model}
 In this section we present a detailed discussion of the coupling of an ensemble of NV centers to a superconducting  transmission line resonator.

\subsection{A. NV spin Hamiltonian}
We start with the Hamiltonian for a single NV center in the electronic ground state in the presence of static and oscillating magnetic fields, $\vec B(t)$. In the weak field limit $\mu_Bg_s|\vec B(t)|\ll \hbar D$, where $D\approx 2.88$ GHz is the zero field splitting, $g_s\simeq 2$ and $\mu_B$ is the Bohr magneton, this Hamiltonian is given by~\cite{SuppNVReview}
\begin{equation}\label{eq:HNVfull}
H_{\rm NV} =  \hbar  D S_z^2  +\mu_B g_s \mathcal{B}_z S_z +\mu_B g_s (\mathcal{B}_x(t) S_x +\mathcal{B}_y(t) S_y)  + H_{\rm inh}.
\end{equation}
Here $\vec S$ is the electronic spin operator (in units of $\hbar$) with components $S_{k=x,y,z}= \vec n_{k} \cdot \vec S$ defined with respect to the local coordinate system $(\vec n_x,\vec n_y, \vec n_z)$ of the NV center, where $\vec n_z$ is aligned with its symmetry axis. The first three terms represent the zero field splitting and the coupling of the spin to static and oscillating magnetic fields, respectively. We have introduced the notation $\mathcal{B}_k=\vec n_k \cdot \vec B$ to distinguish the magnetic field components in the NV center fixed coordinate system  from the components $B_k = \vec e_k\cdot \vec B $ in the laboratory coordinate system $(\vec e_x,\vec e_y, \vec e_z)$. Finally, the last term in Eq.~\eqref{eq:HNVfull} is given by
\begin{equation}
H_{\rm inh} = H_{\rm strain}+ H_{\rm spin} + H_{\rm hyp},
\end{equation}
and accounts for the effect of strain, interactions with neighboring impurity spins and hyperfine interactions, respectively. As discussed in more detail below, these terms lead to random frequency shifts and therefore to an inhomogeneous broadening of the spin ensemble.

\subsubsection{1. Static bias field}
\begin{figure}
\begin{center}
\includegraphics[width=0.70\textwidth]{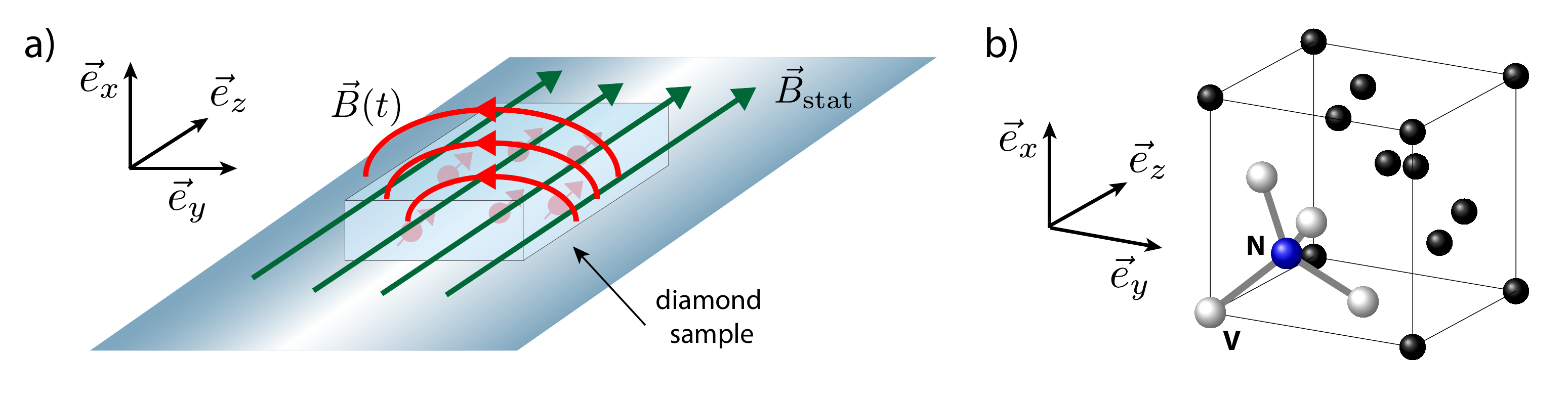}
\caption{a) Orientation of the static bias field $\vec B_{\rm stat}$ (green arrows) and the oscillating microwave fields $\vec B(t)$ (red arrows) relative to the diamond sample. b) The four possible NV center orientations are shown for a diamond cut along the (001) plane (N = nitrogen atom, V = vacancy). With respect to the coordinate system $(\vec e_x,\vec e_y,\vec e_z)$, the four possible orientations of the NV symmetry axis (N-V axis) are $\vec n_z=(1,1,1)/\sqrt{3}$, $\vec n_z=(-1,-1,1)/\sqrt{3}$, $\vec n_z=(1,-1,-1)/\sqrt{3}$ and $\vec n_z=(-1,1,-1)/\sqrt{3}$.}
\label{fig:FieldOrientation}
\end{center}
\end{figure}

The NV centers can have four possible orientations within the diamond lattice and in general the external bias field will affect each of the four groups of NV centers differently. In the following we will consider the special case, where the diamond sample is cut along the (001) plane, and the bias field is applied parallel to this plane along the $\vec e_z$ axis, as shown in Fig.~\ref{fig:FieldOrientation} a).  In this case the projection $\vec e_{z}\cdot  \vec n_z = \pm  1/\sqrt{3}$ is the same for all NV centers, and for a static bias field $\vec B_{\rm stat}=\vec e_zB_{\rm stat} $ the resulting Zeeman splitting $\delta_B$ between the $|\pm 1\rangle$ states is $\delta_B= 2 \mu_B g_s B_{\rm stat}/(\sqrt{3}\hbar)$. Splittings of up to $\delta_B\approx 100$ MHz considered in the main text correspond to applied fields of $B_{\rm stat}<50$ Gauss, which is compatible with superconducting circuits. Note that in Eq.~\eqref{eq:HNVfull} we have neglected the coupling of static fields to the $S_x$ and $S_y$ spin components. Due to the large zero field splitting this coupling can only induce higher order corrections $\sim \delta_B^2/D$, and as long as these shifts are the same for all spins, we can absorb it into a redefinition of $\delta_B$.

\subsubsection{2. Microwave fields}
Near-resonant microwave fields with frequencies $\omega_{\rm mw}  \sim D \sim 3 $ GHz couple to the spin components $S_x$ and $S_y$ orthogonal to the NV symmetry axis. For concreteness, we assume that all microwave fields (classical and quantum) are orthogonal to the static bias field [see Fig.~\ref{fig:FieldOrientation} a)].  For an oscillating field of the form $\vec B(t)= (\vec e_x B_x + \vec e_y B_y)\cos(\omega_{\rm mw} t+\phi_{\rm mw})$ we obtain
\begin{equation}
\begin{split}
H_{\rm mw}= &\mu_B g_s (\mathcal{B}_x S_x +\mathcal{B}_y S_y) \cos(\omega_{\rm mw} t+\phi_{\rm mw})\\
\simeq &\frac{\hbar \Omega_{\rm mw}}{2}\left[  e^{-i(\omega_{\rm mw}t+\phi_{\rm mw})}  \left( e^{-i\theta_{\rm B}} |+1\rangle\langle 0|+ e^{i\theta_{\rm B}} |-1\rangle \langle 0| \right) + {\rm H.c.}\right] + \dots, 
\end{split}
\end{equation}
where the dots represent the remaining off-resonant terms, which can be omitted by making a rotating wave approximation with respect to the large frequency $\omega_{\rm mw}\sim D\gg \Omega_{\rm mw}$.  Here $\Omega_{\rm mw}$ denotes the Rabi frequency of the driving field,
\begin{equation}\label{eq:Rabi_freq}
\Omega_{\rm mw}= \frac{\mu_B g_s}{\sqrt{3}\hbar} \sqrt{B_x^2+B_y^2 \pm B_xB_y},\qquad \tan(\theta_{\rm B})= \mathcal{B}_y/\mathcal{B}_x.
\end{equation}
The $\pm$ sign in Eq.~\eqref{eq:Rabi_freq} depends on the NV orientation, i.e. for two groups of NV centers it will be $+$ and for the the other two groups it will be $-$. As described in the main text, we are interested in the case where the NV centers are driven by two classical microwave fields
\begin{equation}
\vec B_{n=1,2}(\vec r,t)=  \left( B_{n,x}(\vec r)\vec e_x+  B_{n,y}(\vec r)\vec e_y\right) \cos(\omega_{n} t+\phi_{n}).
\end{equation}
The Hamiltonian for the $i$-th NV center located at position $\vec r_i$ in the sample is then given by
\begin{equation}
\begin{split}
H_{\rm NV}^i = &\hbar \left(D- \frac{\delta_B}{2}\right) |-1\rangle_i\langle -1| + \hbar \left(D+  \frac{\delta_B}{2}\right) |+1\rangle_i\langle +1|\\
&+ \sum_{n=1,2} \frac{\hbar \Omega_n^i}{2}  e^{i(\omega_n t+\phi_n)} \left[  \left( e^{i\theta_n^i} |0\rangle_i\langle +1|+ e^{-i\theta_n^i} |0\rangle_i\langle -1|\right)   +   {\rm H.c.}\right]  +  H^i_{\rm inh},
\end{split}
\end{equation}
where the Rabi frequencies $\Omega_n^i$ and field angles $\theta_n^i$ are defined as in Eq.~\eqref{eq:Rabi_freq} above with $B_x\rightarrow B_{n,x}(\vec r_i)$, etc. 

\subsubsection{3. Inhomogeneous broadening of the NV ensemble}\label{sec:Model_Inhomo}
The presence of local strain as well as  the coupling of the NV spin to other electronic or nuclear spins in the surrounding can substantially modify the NV spin levels. Since in the present case the $|\pm 1\rangle$ states are split by a static bias field and a direct coupling of the two states is suppressed, the main effect of these interactions can be accounted for by a random frequency splitting for each NV center,
\begin{equation}
H_{\rm inh}^i \simeq   \frac{\hbar \delta_i }{2} \left( |+1\rangle_i \langle+1|-|-1\rangle_i\langle -1|\right).
\end{equation}
The overall frequency shift, $\delta_i=\delta_i^{\rm strain} + \delta_i^{\rm spin} +\delta_i^{\rm nuc}$,  contains contributions from the strain field, electronic spins and nuclear spins.

\paragraph{Strain.}
Local strain in the diamond lattice breaks the $C_{3v}$ symmetry of the center and modifies the spin level structure. The effect of  strain on the NV center spin states is described by the Hamiltonian~\cite{SuppNVReview,SuppDoldeNatPhys2011}
\begin{equation}
\begin{split}
H_{\rm strain}&=\hbar \gamma_\parallel \mathcal{E}_z S_z^2 +   \hbar \gamma_\perp \mathcal{E}_x  \left(S_x^2 - S_y^2\right) +  \hbar \gamma_\perp \mathcal{E}_y  \left(S_xS_y + S_yS_x\right) \\
&= \hbar \gamma_\parallel \mathcal{E}_z (|+1\rangle\langle +1| + |-1\rangle\langle -1| ) +\hbar \gamma_\perp\left[  (\mathcal{E}_x-i\mathcal{E}_y) |+1\rangle\langle -1| + (\mathcal{E}_x+i\mathcal{E}_y) |-1\rangle\langle +1|\right].
\end{split}
\end{equation}
Here $\mathcal{E}_k$ are the components (in the NV center fixed coordinate system) of the local strain field, and $\gamma_\parallel$ and $\gamma_\perp$ are frequency shifts per unit of strain. In the absence of a magnetic bias field, the transverse coupling leads to a frequency splitting of the $|\pm 1\rangle$ - manifold by $2\gamma_\perp |\mathcal{E}_x+i\mathcal{E}_y|$. Typical values for $\gamma_\perp |\vec{\mathcal{E}}|$ are around a few MHz~\cite{SuppAcostaPRL2010}. However, in the presence of a magnetic bias, this coupling is suppressed and only induces a second order shift of
\begin{equation}
\delta^{\rm strain} = \frac{ 2\gamma_\perp^2 |\mathcal{E}_x+i\mathcal{E}_y|^2}{\delta_B}\lesssim {\rm MHz},
\end{equation}
between the $|\pm 1\rangle$ states. The strain component parallel to the NV axis will lead to a linear, but common shift of the $|\pm1 \rangle$ states, which only weakly influences the two-photon Raman coupling strength, but not the relative detuning between the two excited spin states. Therefore, this parallel strain shift is less important for the current proposal and is neglected in the following.

\paragraph{Spin-spin interactions.}
 The NV center spins will interact with other paramagnetic impurities such as nitrogen atoms with an unpaired electronic spin $S=1/2$ (for current samples with high NV densities typically  only 10-20\% of the nitrogen atoms are converted into NV impurities). 
 The coupling of a single NV center to the surrounding impurity spins is given by \cite{SuppTaylorNatPhys2008,SuppMarcosPRL2010}
\begin{equation}
H_{\rm spin}=\hbar \sum_j  S_z {\bf D}^z_{j} \vec S_j(t).
\end{equation}
Here $\vec S_j$ is the spin operator of the $j$-th impurity spin located at a distance $\vec r_j$ away from the NV center and
\begin{equation}
{\bf D}^z_j= \frac{\mu_0 g_s^2 \mu_B^2}{4\pi\hbar}\frac{3(\vec n_{j}\cdot \vec n_z) \vec n_j - \vec n_z}{|\vec r_j|^3},
\end{equation}
where $\vec n_j = \vec r_j/|\vec r_j|$. For an estimate of the typical splitting $\delta^{\rm spin}$ due to spin-spin interactions, we can simply take twice the value of the dipole coupling strength at the mean distance $r_0 =(3/(4\pi n_{N}))^{1/3}$, where $n_N$ is the density of nitrogen spins. We obtain
\begin{equation}
\delta^{\rm spin}= \frac{2\mu_0 g_s^2 \mu_B^2}{4\pi\hbar r_0^3} S.
\end{equation}
For $S=1/2$ and $n_{N}=10^{19}$ cm$^{-3}$  we obtain values of about $\delta^{\rm spin} \approx 2.3$ MHz.

\paragraph{Hyperfine interactions.}
 Apart from interactions with other electronic spins, the NV spin is affected by hyperfine interactions with nearby nuclear spins. The naturally dominant $^{14}N$ isotope (natural abundance $\sim 99.6\%$) has a nuclear spin $I=1$. In addition, $^{13}C$ atoms in the diamond lattice with natural abundance of $\sim 1.1\%$ have a nuclear spin $I=1/2$. The resulting hyperfine coupling is~\cite{SuppDreauPRB2012}
\begin{equation}
H_{\rm hyp} =  \vec S \mathcal{A}_{N}\vec I_N+ \sum_j \vec S \mathcal{A}^j_{C}\vec I_C^j,
\end{equation}
where $\vec I_N$ is the operator of the nitrogen nuclear spin, the $\vec I_C^j$ denote the operators for the surrounding carbon spins and $\mathcal{A}_{N}$ and $\mathcal{A}_{C}^j$ are the corresponding hyperfine tensors. Under a secular approximation the coupling to  $S_x$ and $S_y$ can be neglected, and assuming that the nuclear spins are static over the relevant timescales we can approximately write
\begin{equation}
H_{\rm hyp} \simeq \frac{\hbar \delta^{\rm nuc}}{2}S_z,\qquad \delta^{\rm nuc}= 2\left(A_N m_N + \sum_j m_{C}^j A_C^j\right).
\end{equation}
The coupling to the $^{14}N$ nuclear spin is well characterized and leads to splitting of the $|0\rangle\leftrightarrow |+1\rangle$ ESR line into three lines $m_N=0,\pm1$ separated by $A_N\approx 2.16 $ MHz~\cite{SuppFeltonPRB2009,SuppDreauPRB2012}. This corresponds to a relative splitting of $4.3$ MHz between the excited spin states. The couplings $A_C^j$ to the $^{13}C$ nuclear spins with $m_C^j=\pm 1/2$ depends strongly on the position of the $^{13}C$ atom in the diamond lattice. A $^{13}C$ atom located directly next to the vacancy results in a hyperfine shift of the $|\pm 1\rangle$ states of $A_C\sim 130$ MHz~\cite{SuppFeltonPRB2009}. Therefore, approximately 3\% of the NV centers are shifted far out of resonance and do not play a role in our model. The hyperfine shifts due to $^{13}C$ atoms on the other lattice sites have been investigated in two recent experimental works~\cite{SuppSmeltzerNJP2011,SuppDreauPRB2012}.  In Ref.~\cite{SuppDreauPRB2012} the ESR splitting of 400 defect centers has been studied. The authors found that roughly 75\% of the centers only show the $^{14}$N splitting and only $\sim$ 15\% exhibit a $^{13}$C splitting  which is larger than 2 MHz, with values up to $14$ MHz. Note that all these values refer to the splitting of the $|0\rangle\leftrightarrow |+1\rangle$ ESR line, and the corresponding values of the excited state splitting $\delta^{\rm nuc}$ are by a factor of 2 larger.

\paragraph{Summary.}
From the above estimates we expect the inhomogeneous frequency distribution $\delta_i$ of a dense NV center ensemble to consist of three main hyperfine peaks at $(0,\pm1)\times 4.4$ MHz, which are smeared out by a couple of MHz by spin-spin interactions and residual strain induced shifts. Current experiments with high density NV samples are rather consistent with a smooth frequency distribution with an inhomogeneous linewidth (FWHM) of about $\gamma_s \sim20$ MHz \cite{SuppSandnerPRA2012,SuppPutz2014} and the $^{14}N$-splitting is not resolved. This broad distribution is most likely due to a higher density of impurity spins than assumed in our estimates and can probably be  improved with better sample preparation techniques. In Sec.~{\bf IV}Ê
below we present more details on the frequency distributions assumed for the numerical calculations in the main text.

\subsection{B. NV ensemble coupled to a microwave cavity}
In our proposal the diamond sample is placed above a superconducting stripline cavity and in addition to the externally applied classical magnetic fields, the NV centers spins will also couple to the quantized field of the microwave resonator. The total Hamiltonian for this system is
\begin{equation}
H= \hbar \omega_c a^\dag a + \sum_i H_{\rm NV}^i(t)  +H_{\rm int},
\end{equation}
where  $H_{\rm int}$ accounts for the magnetic coupling between the NV centers and the cavity field.
The quantized magnetic field associated with the cavity mode is
\begin{equation}
\vec B(\vec r) = \vec B_0(\vec r)  (a+a^\dag),
\end{equation}
where $\vec B_0(\vec r)$ is the magnetic field distribution per microwave photon. By assuming that also $\vec B_0(\vec r) \perp \vec e_z$ and making a rotating wave approximation with respect to $\omega_c\sim D$ we obtain
\begin{equation}
H_{\rm int}= \sum_i \hbar g_0^i    \left[a^\dag \left( e^{i\theta_0^i} |0\rangle_i\langle +1|+ e^{-i\theta_0^i} |0\rangle_i\langle -1|\right)   +   {\rm H.c.}\right],
\end{equation}
where in analogy to the classical fields we have introduced the couplings 
\begin{equation}
g_0^i= \frac{\mu_B g_s}{\sqrt{3}\hbar}¨º \sqrt{B_{0,x}^2(\vec r_i) +B_{0,y}^2(\vec r_i)\pm B_{0,x}(\vec r_i)B_{0,y}(\vec r_i)},
\end{equation}
and the phases $\tan(\theta_0^i)= \mathcal{B}_{0,y}(\vec r_i)/\mathcal{B}_{0,x}(\vec r_i)$. All together, the full Hamiltonian for the coupled NV ensemble - cavity system reads $(\hbar=1)$
\begin{equation}\label{eq:Hfull_supp}
\begin{split}
H=&  \omega_c a^\dag a + \sum_{i}  \left(D- \frac{\delta^i_B}{2}\right) |-1\rangle_i\langle -1| + \left(D+  \frac{\delta^i_B}{2}\right) |+1\rangle_i\langle +1| \\
&+\sum_{i}  \sum_{n=1,2}  \frac{\Omega_n^i}{2}  \left[  e^{i(\omega_n t+\phi_n)} \left( e^{i\theta_n^i} |0\rangle_i\langle +1|¨º+ e^{-i\theta_n^i} |0\rangle_i\langle -1|\right)  +   {\rm H.c.}\right]\\
&+\sum_{i}  g_0^i    \left[a^\dag \left( e^{i\theta_0^i} |0\rangle_i\langle +1|+ e^{-i\theta_0^i} |0\rangle_i\langle -1|\right)   +   {\rm H.c.}\right],
\end{split}
\end{equation}
where the frequency splittings $\delta_B^i = \delta_B +   \delta_i$ include the random frequency offsets $\delta_i$ discussed above.

%
%

\section{II. Effective Dicke Hamiltonian}\label{sec:AdiabaticElimination}   

This section details the derivation of the effective Hamiltonian given in Eq. (2) of the main text. Our starting point is Hamiltonian~\eqref{eq:Hfull_supp}, which contains two driving fields with different frequency, so the time dependence cannot be simply eliminated by changing to a rotating frame. To perform a systematic perturbation theory we convert the time-dependent Hamiltonian into a time-independent Hamiltonian by formally replacing the two classical fields by two additional cavities with operators $a_1$ and $a_2$ and frequencies $\omega_{1}$ and $\omega_2$, respectively. Then
\begin{equation} \label{eq:H_time_indep}
\begin{split}
H=& \sum_{n=0,1,2} \omega_n a^\dag_n a_n +  \sum_i \left(D-\frac{\delta^i_B}{2}\right) |-1\rangle_i\langle -1| + \left(D+\frac{\delta^i_B}{2}\right)  |+1\rangle_i\langle +1| \\
&+ \sum_{n=0,1,2} \sum_i  {\rm g}_n^i \left( a_n^\dag e^{i\theta_n^i}  |0\rangle_i\langle +1| + a_n e^{-i\theta_n^i} |+1\rangle_i\langle 0 |+ a_n^\dag e^{-i\theta_n^i}|0\rangle_i\langle -1|   +   a_n e^{i\theta_n^i} |-1\rangle_i\langle 0|\right),
\end{split}
\end{equation}
where for a notational purpose  we have set $a_0\equiv a$,  $\omega_0\equiv \omega_c$ and ${\rm g}_0^i\equiv g_0^i$. After performing the perturbation theory, the original system can be recovered by assuming that the modes $a_1$ and $a_2$ are each prepared in a large coherent state $|\alpha_{1,2}(t)\rangle$, such that
\begin{equation}
{\rm g}_n^i\langle \alpha_i(t) | a | \alpha_i(t) \rangle = \frac{\Omega_n^i}{2} e^{-i(\omega_n t+\phi_n)},
\end{equation}
and at the same time taking the limit ${\rm g}_{1,2}^i\rightarrow 0$.

We write the time independent Hamiltonian~\eqref{eq:H_time_indep} as $H=H_0 +H_g$, where
\begin{equation}
H_0= \sum_{n=0,1,2} \omega_n a^\dag_n a_n +  \sum_i \omega_- ^i  |-1\rangle_i\langle -1| + \omega_+ ^i  |+1\rangle_i\langle +1|,
\end{equation}
and $\omega_\pm^i=D\pm \delta_B^i/2$. The coupling term is
\begin{equation}
H_g= \sum_{n=0,1,2}\sum_i  {\rm g}_n^i \left( a_n^\dag e^{i\theta_n^i}  |0\rangle_i\langle +1| + a_n e^{-i\theta_n^i} |+1\rangle_i\langle 0 |+ a_n^\dag e^{-i\theta_n^i}|0\rangle_i\langle -1|   +   a_n e^{i\theta_n^i} |-1\rangle_i\langle 0|\right).
\end{equation}
We then use a Schrieffer-Wolff transformation to eliminate the linear coupling $H_g$ and derive an effective second order Hamiltonian for the $|\pm 1\rangle$ manifold.
The transformation is given by
\begin{equation}
\tilde H= UH U^\dag, \qquad U=e^{iS},
\end{equation}
and
\begin{equation}
\tilde H=  H_0 + H_g +i [S,H_0] + i [S,H_g] -\frac{1}{2}[S,[S,H_0]] + ...
\end{equation}
We choose $i[S,H_0]=-H_g$, such that
\begin{equation}
\tilde H=  H_0 + \frac{i}{2} [S,H_g] + \mathcal{O}(g^3).
\end{equation}
For the operator $S$ we make the ansatz
\begin{equation}
S=  i  \sum_{n=0,1,2}\sum_i      \left[ a_n^\dag  \left( \alpha_{n,+}^i  |0\rangle_i\langle +1 | +  \alpha_{n,-}^i  |0\rangle_i\langle -1 | \right)  -   {\rm H.c.}\right].
\end{equation}
The condition $i[S,H_0]=-H_g$ is satisfied by 
\begin{equation}
\alpha_{n,\pm}^i = \frac{{\rm g}_n^i }{(\omega_\pm^i-\omega_{n})} e^{\pm i\theta_n^i}.
\end{equation}
To evaluate the remaining commutator we group operators as $S=i\sum_n (a_n^\dag B_n - a_n B_n^\dag)$  and $H_g=  \sum_n (a_n C^\dag_n + a_n^\dag C_n)$. Then
\begin{equation}
\begin{split}
\frac{i}{2}[S,H_g]= &\frac{1}{2}\sum_{n,m}\left(  \left[a_n B_n^\dag, a_m^\dag C_m\right] + {\rm H.c.}¨º  \right) \\
 = &\frac{1}{2}\sum_{n,m} \left(  a_n a_m^\dag \left[B_n^\dag, C_m\right] + \delta_{n,m}¨º C_m  B_n^\dag +  {\rm H.c.}¨º  \right).
\end{split}
\end{equation}
In our final model we are only interested in the $|\pm 1\rangle$ subspace and we can omit all terms in the effective Hamiltonian which affect the ground state $|0\rangle$. By denoting by $\mathbbm{P}_1$ the projector on the $m_s=\pm1$ subspace we obtain  $\mathbbm{P}_1C_m  B_n^\dag    \mathbbm{P}_1=0$ and
\begin{equation}
\begin{split}
\mathbbm{P}_1\left[B_n^\dag, C_m\right]\mathbbm{P}_1 = \sum_i\Big(  &(\alpha_{n,+}^i)^* {\rm g}_m^i e^{i\theta_m^i}  |+1\rangle_i\langle +1| + (\alpha_{n,-}^i)^* {\rm g}_m^i e^{-i\theta_m^i}|-1\rangle_i\langle -1| \\
 &+ (\alpha_{n,+}^i)^* {\rm g}_m^i e^{-i\theta_m^i} |+1\rangle_i\langle -1|+(\alpha_{n,-}^i)^* {\rm g}_m^i e^{i\theta_m^i} |-1\rangle_i\langle +1| \Big).
\end{split}
\end{equation}
For a further simplification of the final result we use the fact that in the configuration of interest all the cavity frequencies are far detuned from each other and also the states $|\pm 1\rangle$ are detuned by a large frequency offset $\delta_B$. This allows us to eliminate all energy non-conserving terms. Specifically, we consider a configuration, where the frequencies are approximately tuned to $\omega_0\approx D$ and $\omega_1\approx D+\delta_B$, $\omega_2\approx D-\delta_B$. In this case we obtain
\begin{equation}
H_{\rm eff} =\mathbbm{P}_1\left( H_0 + \frac{i}{2} [S,H_g]\right) \mathbbm{P}_1 =H_0+  H_{\rm stark}+ H_{\rm int},
\end{equation}
where
\begin{equation}
\begin{split}
H_{\rm stark} =\sum_{n} \left(  a^\dag_n a_n+ 1\right) \sum_i  \left(  (\alpha_{n,+}^i)^* {\rm g}_n^i e^{i\theta_n^i} |+1\rangle_i\langle +1|¨º + (\alpha_{n,-}^i)^* {\rm g}_n^i e^{-i\theta_n^i} |-1\rangle_i\langle -1|\right),
\end{split}
\end{equation}
and
\begin{equation}
    H_{\rm int}=\sum_i \tilde g_1^i\left(a_0^\dag a_1 e^{-i(\theta_1^i+\theta_0^i)} |+1\rangle_i\langle -1|+{\rm H.c.}\right)+\tilde g_2^i\left(a_0^\dag a_2 e^{i(\theta_2^i+\theta_0^i)} |-1\rangle_i\langle +1|+{\rm           H.c.}\right).
 \end{equation}
In this last term we have introduced the effective Raman couplings
\begin{equation}
\tilde g^i_{1}= \frac{1}{2}\left( \frac{{\rm g}_{1}^i {\rm g}_0^i}{\omega_-^i-\omega_0}+  \frac{{\rm g}_{1}^i {\rm g}_0^i}{\omega_+^i-\omega_1 }  \right),\qquad \tilde g^i_{2}= \frac{1}{2}\left( \frac{{\rm g}_{2}^i {\rm g}_0^i}{\omega_+^i-\omega_0}+  \frac{{\rm g}_{2}^i {\rm g}_0^i}{\omega_-^i-\omega_2 }  \right).
\end{equation}
We now replace the modes $a_1$ and $a_2$ by their classical mean values, ${\rm g}_n^i a_n\rightarrow  \Omega_n^i e^{-i(\omega_n t+\phi_n)}/2$, as described above. Assuming $\omega^i_\pm\approx D\pm \delta_B/2$, $\omega_0\approx D$ and $\omega_1\approx D+\delta_B$, $\omega_2\approx D-\delta_B$, we obtain
\begin{equation}
\begin{split}
H_{\rm stark} =\sum_i & \left(  \Delta_+^i |+1\rangle_i\langle +1|¨º + \Delta_-^i |-1\rangle_i\langle -1|\right)+  \sum_i  \lambda_i \left(a^\dag a +1\right)\left( |+1\rangle_i\langle +1|- |-1\rangle_i\langle -1|\right),
 \end{split}
\end{equation}
where
\begin{equation}
\Delta_+^i =\frac{|\Omega_2^i|^2}{6\delta_B}- \frac{|\Omega_1^i|^2}{2\delta_B},\qquad \Delta_-^i =\frac{|\Omega_2^i|^2}{2\delta_B}- \frac{|\Omega_1^i|^2}{6\delta_B},\qquad \lambda_i=\frac{2(g_0^i)^2}{\delta_B}.
\end{equation}
For the coupling term we obtain
\begin{equation}
\begin{split}
H_{\rm int} =  \sum_i  & g^i_1 \left(  a_0 e^{i(\omega_1 t+\phi_1)}e^{+i(\theta_1^i+\theta_0^i)}  |-1 \rangle_i\langle +1|  + a_0^\dag  e^{-i(\omega_1 t+\phi_1)} e^{-i(\theta_1^i+\theta_0^i)}  |+ 1\rangle_i\langle -1| \right)\\
& +   g^i_2 \left(  a_0 e^{i(\omega_2 t+\phi_2)} e^{-i(\theta_2^i+\theta_0^i)} |+1\rangle_i\langle -1|  +a_0^\dag  e^{-i(\omega_2 t+\phi_2)} e^{i(\theta_2^i+\theta_0^i)} |-1\rangle_i\langle +1| \right),
\end{split}
\end{equation}
where now
\begin{equation}
g^i_1=\frac{g_0^i\Omega_1^i}{4}\left(  \frac{1}{\omega_-^i-\omega_0 } + \frac{1}{\omega_+^i-\omega_1 }   \right) \simeq -  \frac{g_0^i\Omega_1^i}{\delta_B},\qquad g^i_2=\frac{g_0^i\Omega_2^i}{4}\left(  \frac{1}{\omega_+^i-\omega_0 } + \frac{1}{\omega_-^i-\omega_2 }   \right) \simeq + \frac{g_0^i\Omega_2^i}{\delta_B}.
\end{equation}
For concreteness we set $\phi_1=\pi$ and $\phi_2=0$. Further, by assuming that the two classical microwave fields have the same field distribution, we have $\theta_1^i=\theta_2^i$, and in this case also all the $\theta_n^i$ can be absorbed by redefining $e^{-i(\theta_1^i+\theta_0^i)}|+1\rangle_i \rightarrow |+1\rangle_i$.
Then, all together we obtain the effective Hamiltonian
\begin{equation}
\begin{split}
H_{\rm eff}= \omega_c a^\dag a +  &\sum_i  \left[  \frac{(\omega_+^i- \omega^i_-+\Delta_+^i-\Delta_-^i )}{2}+ \lambda_i   a^\dag a \right] \left( |+1\rangle_i\langle +1| -  |-1\rangle_i\langle -1|\right)  \\
&+ \sum_i   g_1^i \left(  a_0 e^{i\omega_1 t} |-1 \rangle_i\langle +1|  + a_0^\dag  e^{-i\omega_1 t} |+ 1\rangle_i\langle -1|\right) + g_2^i \left( a_0 e^{i\omega_2 t}  |+1\rangle_i\langle -1|  +a_0^\dag  e^{-i\omega_2 t} |-1\rangle_i\langle +1| \right),
\end{split}
\end{equation}
where we have omitted an overall common shift of the excited spin states.
In a final step we move into a rotating frame with respect to
\begin{equation}
H=\left(\frac{\omega_1+\omega_2}{2}\right)  a^\dag a +  \left(\frac{\omega_1-\omega_2}{2}\right) \sum_i \left( |+1\rangle_i\langle +1| -  |-1\rangle_i\langle -1|\right),
\end{equation}
and obtain
\begin{equation}
\begin{split}
H_{\rm eff}=& \Delta_c a^\dag a +  \sum_i  \left[  \frac{\Delta^i_s}{2}+ \lambda_i  a^\dag a \right] (\sigma_z^i +1)+ \sum_i  g_1^i \left(  a  \sigma_-^i +a^\dag\sigma_+^i\right) + g_2^i \left(  a  \sigma_+^i +a^\dag\sigma_-^i\right).
\end{split}
\end{equation}
Here we have defined effective cavity and spin frequencies
\begin{equation}\label{eq:EffectiveFreq}
\Delta_c = \omega_c - \left(\frac{\omega_1+\omega_2}{2}\right)- \sum_i \lambda_i , \qquad \Delta_s^i = \delta_B -\left(\frac{\omega_1-\omega_2}{2}\right)+\delta_i +(\Delta_+^i-\Delta_-^i)  - \frac{|\Omega_2^i|^2}{3\delta_B}- \frac{|\Omega_1^i|^2}{3\delta_B},
\end{equation}
which can be adjusted by detuning the microwave fields slightly from the exact two-photon resonance condition.

%
%

\section{III. Superradiant phase transition of the inhomogeneous Dicke model}\label{sec:InhomoDM}
By assuming $g_1^i=g_2^i= g_i$ the effective model derived in the previous section and given in Eq. (2) in the main part of the paper is
 \begin{equation}\label{eq:DMgeneral_Supp}
H_{\rm eff}= \Delta_c a^\dag a +  \sum_i  \left[  \frac{\Delta_s^i}{2}+ \lambda_i  a^\dag a \right] (\sigma_z^i+1) + \sum_i  g_i \left(  a  + a^\dag\right)  \sigma_x^i.
\end{equation}
The full system dynamics is described by a master equation
\begin{equation}
\dot \rho = -i [H_{\rm DM}, \rho] + \kappa (2a\rho a^\dag - a^\dag a \rho - \rho a^\dag a),
\end{equation}
where $2\kappa$ is the photon loss rate. 

\subsection{A. Superradiant transition of the homogeneous DM}
As a reference we first briefly review the non-equilibrium superradiant transition of the standard (homogeneous) DM following closely the analysis presented in Ref.~\cite{SuppDimerPRA2007}. 
For homogeneous couplings, $g_i=g$, and spin frequencies, $\Delta_s^i=\Delta_s$, Eq.~\eqref{eq:DMgeneral_Supp} can be written as
\begin{equation}
H_{\rm DM}= \Delta_c   a^\dag a +  \Delta_s J^z+ 2\lambda  a^\dag a (J^z+\mathcal{N}/2) + \frac{G}{\sqrt{\mathcal{N}}} \left(  a  + a^\dag\right)(J^+ +J^-).
\end{equation} 
Here $\mathcal{N}$ is the total number of spins, $G=g\sqrt{\mathcal{N}}$ is the collective coupling and $J_z$, $J_\pm$ are collective spin $J=\mathcal{N}/2$ operators,
\begin{equation} 
J^z= \frac{1}{2} \sum_i  \sigma_z^i ,\qquad  J^\pm= \sum_i \sigma_\pm^i. 
\end{equation}
In the limit of large  $\mathcal{N}\gg 1$ the properties of the DM are well described by the mean values for the operators $\langle a\rangle$, $\langle J^-\rangle$ and $\langle J^z\rangle$ and small fluctuations around them. In a semiclassical approximation, where all expectation values are factorized, we obtain 
\begin{equation} 
\begin{split}
\frac{d}{dt}\langle a\rangle=-(i\Delta_c+\kappa) \langle a\rangle-i\frac{G}{\sqrt{\mathcal{N}}}(\langle J^-\rangle+\langle J^+\rangle)- i2\lambda (\langle J^z\rangle +\mathcal{N}/2)\langle a\rangle,
\end{split}
\end{equation} 
\begin{equation} 
\frac{d}{dt}\langle J^-\rangle=-i\left(\Delta_s+2\lambda |\langle a\rangle|^2\right) \langle J^-\rangle+i\frac{2G}{\sqrt{\mathcal{N}}}(\langle a\rangle+\langle a^\dag\rangle)\langle J^z\rangle,
\end{equation} 
\begin{equation}
\frac{d}{dt}\langle J^z\rangle=-i\frac{G}{\sqrt{\mathcal{N}}}(\langle a\rangle+\langle a^\dag\rangle)(\langle J^+\rangle-\langle J^-\rangle).
\end{equation} 
For a system with all spins initially prepared in the $|-\rangle$ state, these equations conserve the quantity $\langle J^z\rangle^2+\langle J^+\rangle \langle J^-\rangle=\mathcal{N}^2/4$ and in steady state we have $\langle J^-\rangle =\langle J^+\rangle$ and $2\langle J^z\rangle/\mathcal{N}=-\sqrt{1-4\langle J^-\rangle^2/\mathcal{N}^2}$. By introducing the scaled variables $\alpha =\langle a\rangle /\sqrt{\mathcal{N}}$, $\beta=2\langle J^-\rangle /\mathcal{N}$ and $ \lambda_{\mathcal{N}}= \lambda \mathcal{N}$ we obtain for the remaining equations
\begin{equation}
 \alpha =- \frac{G \beta}{\Delta_c-i \kappa + \lambda_{\mathcal{N}} \left(1-\sqrt{1-\beta^2}\right)},
\end{equation}
and
\begin{equation}
 \left[\Delta_s+2\lambda_{\mathcal{N}}|\alpha|^2 \right] \beta = -2 G (\alpha+\alpha^*)\sqrt{1-\beta^2}.
\end{equation}
For $\lambda_\mathcal{N}=0$ this set of equations has only a trivial solution $\alpha=\beta=0$ for $G<G_{\rm crit}$, where
\begin{equation}\label{eq:Gcrit_Supp}
G_{\rm crit} = \sqrt{ \frac{\Delta_c\Delta_s}{4}Ê\left(1+\frac{\kappa^2}{\Delta_c^2}\right)}, 
\end{equation}
which is the critical coupling of the standard homogeneous Dicke model in the presence of decay~\cite{SuppDimerPRA2007}. 
Above the transition, $G>G_{\rm crit}$, we obtain 
\begin{equation}
\beta = \mp \sqrt{1- \left(\frac{G_{\rm crit}}{G}Ê\right)^4},\qquad \alpha = \pm \frac{G}{\Delta_c-i\kappa} \sqrt{1- \left(\frac{G_{\rm crit}}{G}Ê\right)^4}.
\end{equation} 

The effect of the additional Stark shift term in the DM has been previously considered, e.g., in Ref.~\cite{SuppBhaseenPRA2012}. To show that for $\lambda_\mathcal{N}>0$ this term does not considerably modify the superradiant phase transition, we assume for simplicity $\kappa=0$ and $\Delta_c=\Delta_s$. Then, the stationary value of $\beta$ satisfies
\begin{equation}\label{eq:betaeq}
\beta^2 = \left(\frac{G}{G_{\rm crit}}Ê\right)^4 \frac{\beta^2(1-\beta^2)}{\left[\zeta(\beta^2)+\frac{2\lambda_{\mathcal{N}}}{\Delta_c} \left(\frac{G}{G_{\rm crit}}\right))^2 \frac{\beta^2}{\zeta(\beta^2)}\right]^2}=F(\beta^2),\qquad \zeta(x)=\left[ 1+\frac{\lambda_{\mathcal{N}}}{\Delta_c}\left(1-\sqrt{1-x}\right)\right].
\end{equation} 
According to our assumption $\lambda_{\mathcal{N}} >0$ and $\zeta(\beta^2\rightarrow 0)\simeq 1$. The initial slope of $F(x=\beta^2)$ is still
given by $F'(0)= \left(\frac{G}{G_{\rm crit}}\right)^4$ and for $F'(0)>1$ a non-trivial solution to Eq.~\eqref{eq:betaeq} with $\beta\neq 0$ exists. Therefore, the superradiant transition still occurs at the critical coupling $G_{\rm crit}$, only the values of $\beta$ and $\alpha$ above the transition will be reduced.

In the normal phase $\langle a\rangle=0$ and the collective spin is almost completely polarized, $\langle J_z\rangle \simeq -\mathcal{N}/2$. In the limit of $\mathcal{N}\gg 1$ we can study fluctuation around the classical equilibrium values by using a Holstein-Primakoff approximation, where spin excitations are treated as bosons,  
\begin{equation}
J^z \simeq b^\dag b - \mathcal{N}/2,\qquad      J^-\simeq \sqrt{\mathcal{N}} b,\qquad  [b,b^\dag]=1.
\end{equation}
Then, to lowest order in the fluctuations,
\begin{equation}
H_{\rm DM} \simeq \Delta_s b^\dag b +\Delta_c a^\dagger a+G(b+b^\dag)(a+a^\dag)+ 2\lambda a^\dag a b^\dag b .
\end{equation} 
Since $\lambda/G\sim 1/\sqrt{N}$ the small non-linear correction can be neglected below the transition point. Then, from the remaining  quadratic form of $H_{\rm DM}$ we obtain a closed set of equations
\begin{equation}
\dot{\vec{v}} = {\bf M} \vec v,
\end{equation}
 for the mean values $\vec v=(\langle a\rangle ,\langle a^\dag\rangle ,\langle b\rangle ,\langle b^\dag\rangle )^T$ 
where
\begin{equation}
{\bf M}= \left(
\begin{array}{cccc}
-i\Delta_c -\kappa  & 0 & -i G & -i G   \\
0  &   +i\Delta_c -\kappa & i G & i G   \\
-i G  &  -iG & -i\Delta_s & 0   \\
 i G  &  iG & 0 & +i\Delta_s    \\
\end{array}
\right).
\end{equation}
The eigenvalues $\Lambda$ of this matrix are determined by the solutions of
\begin{equation}\label{eq:EigenvalueEqHomo}
\det \left( {\Lambda\mathbbm{1}-\bf M}\right)= (\Delta_c^2+(\Lambda+\kappa)^2) (\Delta_s^2+\Lambda^2)- 4G^2 \Delta_c \Delta_s  =0.
\end{equation}
As along as ${\rm Re}\{\Lambda\} < 0,\,\forall \Lambda$, all excitations are damped and the normal phase is stable. The superradiant phase appears, when for one eigenvalue ${\rm Re}\{\Lambda\} > 0$. Exactly at the transition point  ${\rm Re}(\Lambda)=0$, for one of the eigenvalues, which we can then write as $\Lambda=i\Lambda_I$, where $\Lambda_I\in \mathbbm{R}$. In this case Eq.~\eqref{eq:EigenvalueEqHomo} reduces to
\begin{equation}
(\Delta_c^2-\Lambda_I^2+\kappa^2+i\Lambda_I\kappa) (\Delta_s^2-\Lambda_I^2)- 4G^2 \Delta_c \Delta_s  =0,
\end{equation}
and from looking at the imaginary part of this result it follows that also $\Lambda_I=0$. Therefore, at the phase transition point one of the eigenvalues is exactly zero and this occurs at a critical coupling $G=G_{\rm crit}$ identified in Eq.~\eqref{eq:Gcrit_Supp} above.

\subsection{B. The superradiant phase transition of the inhomogeneous DM}
Having reviewed the basic properties of the standard DM, we now generalize the above analysis to the present case of interest, where both the spin-cavity couplings $g_i$ as well as the individual spin frequencies $\Delta_i$ are inhomogeneously distributed. Note that generalizations of the DM for a distribution of $g_i$ have been previously studied, for example, in Ref.~\cite{GotoPRA2008}. In the present systems the most interesting aspect comes from the broad distribution of spin frequencies, which means that $\Delta_s^i\approx 0$ or even $\Delta_s^i < 0$ for part of the system.  

We follow the approach outlined in the main part of the paper and group together spins with approximately the same coupling constant $g_i\simeq g_\mu $ and approximately the same frequency $\Delta_s^i\simeq \Delta_\mu$ into a single collective spin with operators
\begin{equation}
J^{z}_\mu= \frac{1}{2} \sum_{i\in \mu} \sigma_i^{z},\qquad   J^{\pm}_\mu = \sum_{i\in\mu} \sigma_i^{\pm},
\end{equation}
where $N_\mu$ is the number of spins within the sub-ensemble $\mu$. Since in total we have approximately $\mathcal{N}\sim 10^{12}-10^{14}$ spins, each sub-ensemble will still contain a lot of spins, $N_\mu\gg1 $. Therefore,  in the normal phase we can make a Holstein-Primakoff approximation for each sub-ensemble separately,
\begin{equation}
J_z^\mu \simeq b_\mu^\dag b_\mu - N_\mu/2,\qquad      J_\mu^-\simeq \sqrt{N_\mu} b_\mu,
\end{equation}
where $[b_\mu, b^\dag_{\mu'}]=\delta_{\mu,\mu'}$. The resulting quadratic Hamiltonian is  then of the form
\begin{equation}
\begin{split}
H_{\rm DM}\simeq \Delta_c a^\dag a +  \sum_\mu  \Delta_\mu b_\mu^\dag b_\mu+ \sum_\mu G_\mu \left(  a  + a^\dag\right)  \left(b_\mu+b_\mu^\dag\right),
\end{split}
\end{equation}
where $G_\mu=g_\mu \sqrt{N_\mu}$ are the collective couplings for each sub-ensemble. As above we can write the corresponding equation of motion for the mean values $\vec v=(\langle a\rangle ,\langle a^\dag\rangle ,\langle b_1\rangle ,\langle b_1^\dag\rangle , \langle b_2\rangle ,\langle b_2^\dag\rangle, \dots)^T$ in a matrix form as $\dot{\vec{v}} = {\bf M} \vec v$,
where
\begin{equation}
{\bf M}= \left(
\begin{array}{ccccccc}
-i\Delta_c -\kappa  & 0 & -i G_1 & -i G_1 & -i G_2 & -i G_2 & \dots \\
0  &   +i\Delta_c -\kappa & i G_1 & i G_1 & iG_2 & iG_2& \dots \\
-i G_1  &  -iG_1 & -i\Delta_1 & 0  & 0 & 0 & \dots \\
 i G_1  &  iG_1 & 0 & +i\Delta_1  & 0 & 0 & \dots \\
-i G_2  &  -iG_2 & 0 & 0  &-i\Delta_2 & 0 & \dots \\
 i G_2  &  iG_2 & 0 & 0  & 0 & +i\Delta_2 & \dots \\
\dots &   \dots &  \dots &  \dots &  \dots &  \dots &  \dots \\
\end{array}
\right).
\end{equation}
The eigenvalues $\Lambda$ of this matrix are determined by the solutions of
\begin{equation} \label{criticalmulti}
\det \left( {\Lambda\mathbbm{1}-\bf M}\right)= (\Delta_c^2+(\Lambda+\kappa)^2) \prod_\mu (\Delta_\mu^2+\Lambda^2)\left[ 1- \sum_\mu \frac{4G_\mu^2 \Delta_c \Delta_\mu }{(\Delta_c^2+(\Lambda+\kappa)^2)(\Delta_\mu^2+\Lambda^2)}\right] =0.
\end{equation}
Again the phase transition occurs at the point where at least for one eigenvalue the real part changes from a negative to a positive value, and as above we can show that this requires that also ${\rm Im}(\Lambda)=0$.  However, in the present case the frequencies $\Delta_i$ (which are effective detunings) can have a broad distribution and can be close to zero or negative and simply setting $\Lambda=0$ can lead to diverging results.
Instead, we consider an eigenvalue with a small negative value $\Lambda=-\epsilon$, and the phase transition point is then determined by taking the limit
\begin{equation}\label{eq:Gcrit:Inhomo}
\lim_{\epsilon\rightarrow 0}  \sum_\mu \frac{4G_\mu^2 \Delta_c \Delta_\mu }{(\Delta_c^2+\kappa^2)(\Delta_\mu^2+\epsilon^2)}=1.
\end{equation}
Physically, $\epsilon$ can also be interpreted as a finite spin decay rate, which is assumed to be much smaller than the other frequency scales.

\subsection{C. Critical coupling for an inhomogeneously broadened spin ensemble}
The expression for the phase transition point given in Eq.~\eqref{eq:Gcrit:Inhomo} is valid for an arbitrary set of collective spin states. In the following we consider the limit where the distribution of couplings and frequencies is sufficiently dense as it is the case for a large ensemble of NV centers. Notice that the result in Eq.~\eqref{eq:Gcrit:Inhomo} does not crucially depend on how we group the spins (as long as the coarse graining is sufficiently fine), and therefore we can formally take the limit where each group contains only a single spin, $N_\mu\rightarrow 1$, and introduce the normalized spectral density~\cite{SuppWesenbergPRA2011}
\begin{equation}
\rho(\omega)= \frac{1}{G^2} \sum_\mu g_\mu^2 \delta(\omega-\Delta_\mu)\simeq \sum_i g_i^2 \delta(\omega-\Delta_s^i),
\end{equation}
where $G= \sqrt{\sum_\mu g_\mu^2}\simeq \sqrt{\sum_i g_i^2}$ is the generalized collective coupling strength. For a sufficiently dense frequency distribution, $\rho(\omega)$ is a continuous function of $\omega$ and the phase transition point is given by
\begin{equation}\label{eq:Gcrit_Pvalue}
\frac{4 G^2}{\Delta_c\bar \Delta_s(1+\kappa^2/\Delta_c^2)}  \times \mathcal{P} \int d\omega\, \frac{\bar \Delta_s}{\omega} \rho(\omega)=1,
\end{equation}
where  $\mathcal{P}$ denotes the Cauchy principal value, and $\bar \Delta_s$ is the central spin frequency. 

Due to the AC-Stark shift corrections of the effective spin frequencies $\Delta_s^i$ defined in Eq.~\eqref{eq:EffectiveFreq}, both the couplings $g_i$ and frequencies $\Delta_s^i$ depend on the position of the NV center, and in general they are correlated. However, under the assumption that the applied classical fields are sufficiently homogeneous over the sample, the common Stark shift can be absorbed into a shift of the central spin frequency 
\begin{equation}
\bar  \Delta_s = \delta_B  -\left(\frac{\omega_1-\omega_2}{2}\right)- \frac{(|\Omega_1^i|^2+|\Omega_2^i|^2)}{3\delta_B},
\end{equation}
and the normalized spectral density simplifies to $\rho(\omega)\equiv P(\omega)=P_\delta(\omega-\bar \Delta_s)$, where $ P_\delta(\omega)$ is the distribution function of the inhomogeneous frequency offsets $\delta_i$. For the example of a Lorentzian frequency distribution\begin{equation}
P_\delta(\omega) = \frac{1}{2\pi} \frac{\gamma_s}{\omega^2+\gamma_s^2/4},
\end{equation}
with a FWHM of $\gamma_s$, we obtain
\begin{equation}
\mathcal{P} \int d\omega\, \frac{\Delta_s}{\omega} \rho(\omega) = \frac{\Delta^2_s}{\Delta_s^2+\gamma_s^2/4}.
\end{equation}
This corresponds to the modified transition point discussed in the main part of the paper. Note that although the principal value  does not strongly depend on the exact shape of $P(\omega)$, it can differ by a factor $\sim 2$ for a sharper, e.g., a Gaussian distribution with the same FWHM $\gamma_s$. This also results in the reduction of the critical coupling strength shown in Fig 2 c) in the main text. 

\subsection{D. Superradiant phase} 
Once the critical coupling condition is met, the normal phase becomes unstable and the system relaxes into a new stationary state. To evaluate the properties of this new phase we consider the coupled equations of motions for the average field 
\begin{equation} 
\begin{split}
\frac{d}{dt}\langle a\rangle=-(i\Delta_c+\kappa) \langle a\rangle-i\sum_\mu \frac{G_\mu}{\sqrt{N_\mu}}(\langle J_\mu^-\rangle+\langle J_\mu^+\rangle)- i2\sum_\mu \lambda_\mu (\langle J_\mu^z\rangle +N_\mu/2)\langle a\rangle,
\end{split}
\end{equation} 
and the average spin components
\begin{equation} 
\frac{d}{dt}\langle J_\mu ^-\rangle=-i\left(\Delta_\mu+2\lambda_\mu |\langle a\rangle|^2\right) \langle J_\mu^-\rangle+i\frac{2G_\mu}{\sqrt{N_\mu}}(\langle a\rangle+\langle a^\dag\rangle)\langle J_\mu^z\rangle,
\end{equation} 
and
\begin{equation}
\frac{d}{dt}\langle J_\mu^z\rangle=-i\frac{G_\mu}{\sqrt{N_\mu}}(\langle a\rangle+\langle a^\dag\rangle)(\langle J_\mu^+\rangle-\langle J_\mu^-\rangle).
\end{equation}
Again, the quantity $\langle J_\mu^z\rangle^2+\langle J_\mu^+\rangle \langle J_\mu^-\rangle=N_\mu^2/4$ is conserved and for an initially fully polarized state we have $2\langle J_\mu^z\rangle/N_\mu=-\sqrt{1-4|\langle J_\mu^-\rangle|^2/N_\mu^2}$. For the remaining expectation values we introduce the scaled variables $\alpha =\langle a\rangle /\sqrt{N_c}$, and $\beta_\mu=2\langle J_\mu^-\rangle /N_\mu$. Here the characteristic cavity photon number $N_c$ is defined by $\sqrt{N_c}=(\sum_\mu G_\mu \sqrt{N_\mu})/G$, and is chosen such that in the limit of fully saturated spins, $\beta_\mu\rightarrow 1$,
\begin{equation}
\frac{\langle a\rangle }{\sqrt{N_c}} \approx  \pm \frac{ G}{(\Delta_c +\sum_\mu \lambda_\mu N_\mu ) -i\kappa}.
\end{equation}
Note that the presence of the Stark shift term also reduces the spin saturation in the regime $G\gg G_{\rm crit}$ and in general the value of $\langle a\rangle$ is slightly smaller. 
With these definitions we obtain
\begin{equation}\label{eq:NormEq_alpha}
\begin{split}
\dot \alpha =-\left(i\Delta_c+i\sum_\mu \lambda_\mu N_\mu (1-\sqrt{1-|\beta_\mu|^2})+\kappa\right) \alpha-i\sum_\mu \frac{G_\mu}{2} \sqrt{\frac{N_\mu}{N_c}}(\beta_\mu+\beta_\mu^*),
\end{split}
\end{equation} 
and
\begin{equation} \label{eq:NormEq_beta}
\dot \beta_\mu=-i\left(\Delta_\mu+2\lambda_\mu N_c |\alpha|^2\right) \beta _\mu-i2G_\mu \sqrt{\frac{N_c}{N_\mu}}(\alpha+\alpha^*)\sqrt{1-|\beta_\mu|^2},
\end{equation} 
and for a given distribution of couplings and frequencies the stationary values of $\alpha$ and $\beta_\mu$ can be obtained numerically.

\section{IV. Coupling and frequency distribution: Example}\label{sec:Parameters} 
To demonstrate the feasibility of our scheme under realistic experimental conditions, we perform a detailed estimate of the distribution of couplings and frequencies expected for typical experimental settings. For clarity we write in this section the group index $\mu$ explicitly as a pair of indices $\mu\rightarrow (\mu,\nu)$, where $\mu$ labels groups of spins with approximately the same coupling and $\nu$ labels different frequencies. Therefore, we divide the whole spin ensemble into sub-ensembles of $N_{\mu,\nu}$ spins with bare coupling $g_0^i\approx g_{0,\mu}$ and frequency shift $\delta_i\approx \delta_\nu$. Under the assumption that the classical fields are sufficiently homogeneous over the sample and $\Omega_n^i\simeq\Omega$, we then define 
\begin{equation}
G_{\mu,\nu}= \frac{g_{0,\mu} \Omega}{\delta_B} \sqrt{N_{\mu,\nu}} ,\qquad \lambda_{\mu,\nu} =Ê\frac{2 g_{0,\mu}^2}{\delta_B}, 
\end{equation}Ê
and the collective coupling and the characteristic cavity photon number are given by
\begin{equation} 
G = \sqrt{\sum_{\mu,\nu}Ê G_{\mu,\nu}^2},\qquad  N_c=\frac{1}{G^2} \left[\sum_\mu g_\mu \left(\sum_\nu N_{\nu,\mu}\right)\right]^2.
\end{equation} 
Since for a homogeneous $\Omega_n^i$ the frequency and the coupling distribution are uncorrelated, the number of spins $N_{\mu,\nu}$ with a given frequency $\Delta_\nu=\bar \Delta+\delta_\nu$ is given by
\begin{equation}
\frac{N_{\mu,\nu}}{(\sum_\nu N_{\nu,\mu})}= P(\Delta_\nu)=P_\delta(\delta_\nu=\Delta_\nu-\bar \Delta_s),
\end{equation}
where $P_\delta$ is the distribution of inhomogeneous frequency offsets $\delta_i$. 
These parameters are then used to integrate the normalized set of coupled equations given in Eqs.~\eqref{eq:NormEq_alpha} and~\eqref{eq:NormEq_beta}. 
For the numerical results shown in the main part of the paper we have used 15 different values for $g_\mu$ and $51$ different values for $\delta_\nu$. The Zeeman splitting is $\delta_B=100$ MHz and the Rabi-frequency $\Omega$ is varied between 0 and 20 MHz. If not stated otherwise the following values for $g_\mu$ and $G$, etc. refer to the maximally achievable couplings at $\Omega=20$ MHz.

\subsection{A. Couplings}
\begin{figure}
\begin{center}
\includegraphics[width=0.95\textwidth]{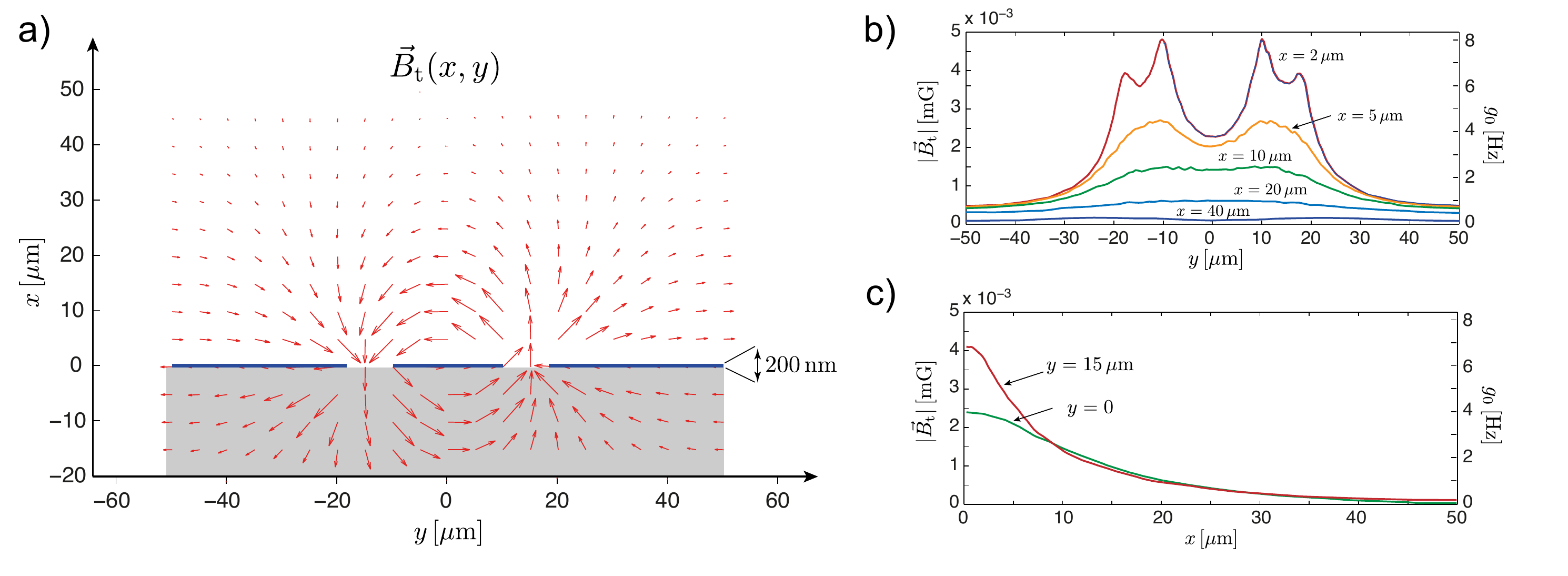}
\caption{a). Distribution of the transverse magnetic field (per photon) above a planar transmission line cavity. b) and c) Value of the magnetic field strength $|B_{\rm t}|$ and the corresponding single spin coupling strength $g_0=\mu_B g_s |B_{\rm t}|/(\sqrt{3}\hbar)$ at various positions above the substrate.
}
\label{fig:Supp_gDistribution}
\end{center}
\end{figure}

As derived in Sec.~{\bf I} 
the bare coupling strength between a single spin and the quantized cavity field is
\begin{equation}
g_0^i= \frac{\mu_B g_s}{\sqrt{3}\hbar}B_{0,i}^\perp,\qquad   B_{0,i}^\perp=\sqrt{B_{0,x}^2(\vec r_i) +B_{0,y}^2(\vec r_i)\pm B_{0,x}(\vec r_i)B_{0,y}(\vec r_i)}.
\end{equation}
For a TEM mode in a transmission line cavity of length $L_c\sim 5$ cm in z-direction, the magnetic field distribution per photon is
\begin{equation}
\vec B_0(\vec r)= \vec B_{\rm t}(x,y) \sin(\pi z/L_c).  
\end{equation} 
The transverse field $\vec B_{\rm t}(x,y)$ is simulated numerically for a typical transmission line geometry and plotted in Fig.~\ref{fig:Supp_gDistribution} a). At a few $\mu$m above the surface the absolute value of the magnetic field is a few $10^{-3}$ milligauss, which corresponds to a bare spin-cavity coupling $g_0$ of a few Hz.  Fig. 2 a) in the main part of the paper shows a histogram of values of the bare spin couplings $g_0^i$ obtained for a diamond sample with dimensions $(l_x,l_y,l_z)=(50,100,500)$ $\mu{\rm m}$, placed on top of the electrodes, and assuming a  density of NV centers of $n_{\rm NV}=10^{18}$ cm$^{-3}$ ($\approx 6$ ppm). 
For this example we obtain the characteristic ensemble quantities 
\begin{equation}
G_0\approx 7.5\,{\rm MHz}, \qquad G\approx 1.5 \, {\rm MHz},\qquad N_c\approx 3.17 \times 10^6,\qquad \sum_\mu \lambda_\mu N_\mu \approx 1.4\,{\rm MHz}. 
\end{equation} 
Note that the cross section of the actual diamond samples used in experiments are much larger than the dimensions assumed here and the value of $G_0$ slightly underestimates the experimentally observed values of $G_0\approx 10$ MHz~\cite{SuppKuboPRL2010,SuppAmsussPRL2011}.

\subsection{B. Frequency distribution}
The dominant sources of inhomogeneous line broadening for an ensemble of NV centers have been discussed in Sec.~{\bf I}. 
 In experiments, the actual width and shape of the spin frequency distribution can be deduced from transmission spectra~\cite{SuppSandnerPRA2012} or dynamical studies~\cite{SuppPutz2014}. In the samples analyzed in Refs.~\cite{SuppSandnerPRA2012,SuppPutz2014}, the reconstructed lineshape for the $|0\rangle\rightarrow |+1\rangle$ transitions is consistent with a $q$-Gaussian distribution with FWHM $\gamma_q^{(0,1)}\approx 10$ MHz and a parameter $q\approx 1.3$. The $q$-Gaussian distribution is defined as
\begin{equation}
P_\delta(\omega)=C_q \left[1-(1-q) \omega^2/a^2 \right]^{1/(1-q)},
\end{equation}
where $C_q$ is a normalization constant. For $1<q\leq 2$ this distribution  interpolates between a Gaussian $(q\rightarrow1)$ and a Lorentzian $(q=2)$ distribution. with a full width at half maximum (FWHM) of $\gamma_q=2 a\sqrt{2^q-2/(2q-2)}$ and $C_q=\sqrt{(q-1)/(\pi a^2)}\Gamma[1/(q-1)]/\Gamma[(3-q)/(2(q-1))]$.
Note that here we are interested in the splitting $\delta_i$ between the $|\pm 1\rangle$ states, and the corresponding width is twice as large, i.e., $\gamma_s=2\gamma_q^{(0,1)}$.  For illustrational purposes and due to a slow numerical convergence, the numerical results presented in the main text are evaluated for slightly smaller values of $\gamma_s$, and the results for a $q$-Gaussian ensemble are compared with the results for a simpler Lorentzian distribution.

\section{V. Non-equilibrium phases of the Dicke lattice model}\label{sec:DickeLatticeModel}
In this section we evaluate the non-equilibrium phase diagram of the Dicke lattice model (DLM), which is shown in Fig. 3 in the main text. The DLM is described by the Hamiltonian
\begin{equation} \label{eq:DickeLattice}
\begin{split}
H_{\rm DLM} &=\Delta_c \sum_{\ell=1}^{N_L} a^\dagger_{\ell} a_{\ell}   -t\sum_{\ell=1}^{N_L-1}(a^\dagger_\ell a_{\ell+1}+a_\ell a^\dag_{\ell+1})+\Delta_s \sum_{\ell=1}^{N_L}   J^z_\ell + \sum_{\ell=1}^{N_L} \frac{G}{\sqrt{\mathcal{N}}}(J^+_{\ell}+J^-_\ell)(a_\ell+a_\ell^\dag),
\end{split}
\end{equation}
where the $a_\ell$ are the bosonic operators for each cavity and the $J_\ell^{\pm,z}$ are collective spin operators for a total spin  $J_\ell=\mathcal{N}/2$. The  system described by the master equation
\begin{equation}
\dot \rho= -i[H_{\rm DLM},\rho] + \kappa \sum_\ell (2a_\ell \rho a^\dag_\ell- a^\dag_\ell a_\ell \rho- \rho a^\dag_\ell
 a_\ell),
\end{equation} 
where an equal photon decay rate $2\kappa$ for cavities has been assumed.
 
\subsection{A. Normal phase} 
For $t,G\rightarrow 0$ the system relaxes into the normal phase, where all cavity modes are in a vacuum state, $\langle a_\ell\rangle=0$, and all spin ensembles are fully polarized. Similar to the case of the single cavity setup, we analyze the stability of the normal phase by making a 
Holstein-Primakoff approximation for each spin ensemble,
\begin{equation}
J_\ell^-=\sqrt{\mathcal{N}}b_\ell,\quad J_\ell^z=b_\ell^\dag b_\ell-\frac{\mathcal{N}}{2},
\end{equation}
 where $[b_\ell,b_{\ell'}^\dag]=\delta_{\ell\ell'}$. Further, by considering for now periodic boundary conditions, we introduce momentum modes
\begin{equation}
a_k = \frac{1}{\sqrt{N_L}} \sum_\ell   e^{i k\ell}a_\ell,\qquad b_k = \frac{1}{\sqrt{N_L}} \sum_\ell   e^{i k\ell}b_\ell,    
\end{equation}
where $k= \pi/N_L\times n$, $n=-N_L+2,N_L+4,\dots,N_L$. Then, the Hamiltonian can be written as
\begin{equation} \label{super1h}
\begin{split}
H_{\rm DLM}=\sum_k\Delta_s b_k^\dagger b_k+\sum_k\Delta_k a^\dagger_k a_k&+G\sum_k(b^\dagger_k a_k+b_k a^\dagger_k + b_k^\dagger a^\dagger_{-k}+b_k a_{-k}),
\end{split}
\end{equation}
where $\Delta_k= \Delta_c-2t\cos(k)$. Similarly,
\begin{equation}
\dot \rho= -i[H_{\rm DLM},\rho] + \kappa \sum_k (2a_k \rho a^\dag_k- a^\dag_k a_k \rho- \rho a^\dag_k
 a_k).
\end{equation}
For each wavevector $k$ we obtain  a closed set of equations of motion for the mean values 
\begin{equation} \label{eom1}
\frac{d}{dt}
\left(
\begin{array}{c}
\langle b_k\rangle\\ \langle a_k\rangle\\ \langle b_{-k}^\dagger\rangle\\ \langle a_{-k}^\dagger\rangle
\end{array}\right)
={\bf M}_k
\left(
\begin{array}{c}
\langle b_k\rangle\\ \langle a_k\rangle\\ \langle b_{-k}^\dagger\rangle\\ \langle a_{-k}^\dagger\rangle
\end{array}\right),
\end{equation}
where
\begin{equation}
{\bf M}_k=
\left(
\begin{array}{cccc}
-i\Delta_s&-i G&0&-i G\\
-i G&-i\Delta_k-\kappa &-i G&0\\
0&i G&i\Delta_s&i G\\
i G&0&i G&i\Delta_{k}-\kappa
\end{array}\right).
\end{equation}
Using similar arguments as in Sec.~{\bf III}, 
it can be shown that for $\Delta_k,\Delta_s>0$, one of the eigenvalues of ${\bf M}_k$ will develop a positive real part when the coupling $G$ exceeds the value
\begin{equation}
G_k=\sqrt{\frac{\Delta_k\Delta_s}{4}\left(1+\frac{\kappa^2}{\Delta_k^2}\right)}.
\end{equation}
Therefore, the phase transition point is given by $G_{\rm crit}={\rm min}\{G_{k}| k \}$. For  $\Delta_c-2t>\kappa $ we obtain
\begin{equation}
G_{\rm crit}=G_{k=0}=\frac{1}{2}\sqrt{\Delta_s(\Delta_c -2t)\left(1+\frac{\kappa^2}{(\Delta_c-2t)^2}\right)},
\end{equation}
and the normal phase becomes unstable due to fluctuations of the homogeneous, $k=0$ mode. 
For the parameter regime $0<\Delta_c-2t<\kappa<\Delta_c+2t$ we obtain instead 
\begin{equation} \label{boundary2}
G_{\rm crit}=G_{k=k_c}=\sqrt{\frac{\kappa\Delta_s}{2}}.
\end{equation}
The critical wavevector $k_c$ is determined by the condition $\Delta_{k_c}=\kappa$, or  $k_c=\arccos((\Delta_c-\kappa)/2t)$. This shows that for finite $\kappa$ the phase transition can be driven by fluctuations with a non-zero wavevector $k_c\neq 0$. For any larger value of $\kappa$ we find that $G_{\rm crit}=G_{k=\pi}$. Finally, as soon as $2t>\Delta_c$ (still assuming $\Delta_s>0$) at least one of the frequencies  $\Delta_k<0$ and one eigenvalue $\Lambda$ of ${\bf M}_k$ has ${\rm Re}\{\Lambda\} >0$ even for $G\rightarrow 0$.

\subsection{B. Superradiant phases} 
Beyond the critical coupling strength the normal phase is unstable and the system relaxes into a new stationary state. The corresponding phases can be characterized by the mean values of the cavity operators $\langle a_\ell\rangle$, which in the limit of large collective spins, $J_\ell\gg1$, can be obtained from the coupled set of semiclassical equations of motions,
\begin{equation} \label{meanfield1}
\begin{split}
\frac{d}{dt}\langle a_\ell\rangle=-(i\Delta_c+\kappa) \langle a_\ell\rangle- i\frac{G}{\sqrt{\mathcal{N}}}(\langle J_\ell^-\rangle+\langle J_\ell^+\rangle)+ i t(\langle a_{\ell+1}\rangle+\langle a_{\ell-1}\rangle),
\end{split}
\end{equation}
\begin{equation} \label{meanfield2}
\frac{d}{dt}\langle J_\ell^-\rangle=-i\Delta_s\langle J_\ell^-\rangle + i 2\frac{G}{\sqrt{\mathcal{N}}}(\langle a_\ell\rangle+\langle a_\ell^\dag\rangle)\langle J_\ell^z\rangle,
\end{equation}
\begin{equation}\label{meanfield3}
\frac{d}{dt}\langle J_\ell^z\rangle=-i\frac{G}{\sqrt{\mathcal{N}}}(\langle a_\ell\rangle+\langle a_\ell^\dag\rangle)(\langle J_\ell^+\rangle-\langle J_\ell^-\rangle).
\end{equation}
Note that these equations are non-linear, and in general complicated dynamics and chaotic behavior can be expected. In the following we focus on damped systems with $\kappa>0$ and not too large values of $G$. In this case the system is reasonably well behaved for $t<\Delta_c/2$, and in this regime we find three types of stationary solutions, which we confirm numerically. The first solutions corresponds to the normal phase described above, the other two type of solutions correspond to homogeneous superradiant phase, and a superradiant phase with broken translational symmetry. 

\subsubsection{1. Superradiant phase: homogeneous}

As discussed above, for small $t$ the $k=0$ mode becomes unstable first, and we can assume that also beyond the phase transition point the system will evolve into a homogeneous phase. By making the ansatz $\langle a_\ell\rangle=\langle a\rangle$ and $\langle J_\ell^{z,\pm}\rangle =\langle J^{z,\pm}\rangle$, we obtain the steady state solutions
\begin{equation}
\langle a\rangle=\pm\frac{G\sqrt{\mathcal{N}}}{\Delta_c-2t-i\kappa}\sqrt{1-\left(\frac{G_{\rm crit}}{G}\right)^4},
\end{equation}
and
\begin{equation}
\langle J^-\rangle=\mp\frac{\mathcal{N}}{2}\sqrt{1-\left(\frac{G_{\rm crit}}{G}\right)^4},\qquad \langle J^z\rangle=-\frac{\mathcal{N}}{2}\left(\frac{G_{\rm crit}}{G}\right)^2,
\end{equation}
which resemble the results obtained for a single cavity, but with a reduced cavity frequency $\Delta_c\rightarrow \Delta_{k=0}=\Delta_c-2t$.
To check the stability of this homogeneous superradiant phase, we perform a fluctuation analysis by introduce approximate bosonic modes $c_\ell$ and $d_\ell$ via~\cite{SuppDimerPRA2007}
\begin{equation}â
a_\ell = \langle a\rangle+c_\ell  \qquad b_\ell = d_\ell+ \frac{\langle J_\ell^-\rangle}{\sqrt{\frac{\mathcal{N}}{2}(1+\nu)}},
\end{equation}
where $\nu=G_{\rm crit}^2/G^2$. Up to the second order of fluctuations and by changing again into $k$-space, the master equation can be written as
\begin{equation}
\dot\rho=-i[H,\rho]+\kappa\sum_k\left(2c_k\rho c_k^\dag-c_k^\dag c_k\rho-\rho c_k^\dag c_k\right)
\end{equation}
with Hamiltonian 
\begin{equation}
\begin{split}
H=&\sum_k \Delta_k c_k^\dag c_k+\Delta_s(\nu) \sum_k d_k^\dag d_k+\mathcal{E}(\nu)   \sum_k (d_k^\dag d_{-k}^\dag+d_{-k}d_k)+G(\nu)\sum(c_kd_k^\dag+c_k^\dag d_k+c_k^\dag d_{-k}^\dag+c_kd_{-k}),
\end{split}
\end{equation}
where we have introduced the abbreviations
\begin{equation}
\begin{split}
\Delta_s(\nu)= \Delta_s \left(\frac{(1+\nu)}{2\nu}+\frac{(1-\nu)(3+\nu)}{4\nu(1+\nu)}\right),\qquad \mathcal{E}(\nu)=\Delta_s\frac{(1-\nu)(3+\nu)}{8\nu(1+\nu)}, \qquad G(\nu)= G\nu\sqrt{\frac{2}{1+\nu}}.
\end{split}
\end{equation}
For each wavevector $k$ we obtain a closed set of equations of motion for the mean values of the fluctuations
\begin{equation}
\frac{d}{dt}
\left(
\begin{array}{c}
\langle d_k\rangle\\ \langle c_k\rangle\\ \langle d_{-k}^\dagger\rangle\\ \langle c_{-k}^\dagger\rangle
\end{array}\right)
={\bf M}_k
\left(
\begin{array}{c}
\langle d_k\rangle\\ \langle c_k\rangle\\ \langle c_{-k}^\dagger\rangle\\ \langle d_{-k}^\dagger\rangle
\end{array}\right),
\end{equation}
where now
\begin{equation}
{\bf M}_k=
\left(
\begin{array}{cccc}
-i\Delta_s(\nu) &-i G(\nu)&-2i\mathcal{E}(\nu)&-i G(\nu)\\
-i G(\nu)&-i\Delta_k-\kappa&-i G(\nu)&0\\
2i\mathcal{E}(\nu)&i G(\nu)&i\Delta_s(\nu)&i G(\nu)\\
i G(\nu)&0&i G(\nu)&i\Delta_k-\kappa
\end{array}\right).
\end{equation}
By checking numerically whether one of the eigenvalues of ${\bf M}_k$ has a positive real part, we get the boundary of the stable region of this homogeneous superradiant phase. For $\Delta_c-2t>\kappa$ this boundary agrees with the phase transition line obtained from the stability analysis of the normal phase. In this regime we also confirm the validity of the homogeneous ansatz by numerically integrating the coupled equations \eqref{meanfield1}-\eqref{meanfield3} for arrays up to $N_L=100$ cavities.
For $0<\Delta_c-2t<\kappa$ the boundary of the homogeneous superradiant phase and the normal phase do no longer coincide.  Therefore, there is a region in the parameter space, where the assumption of a homogeneous phase breaks down.

\subsubsection{2. Superradiant phase at finite momentum}
\begin{figure}
\begin{center}
\includegraphics[width=0.75\textwidth]{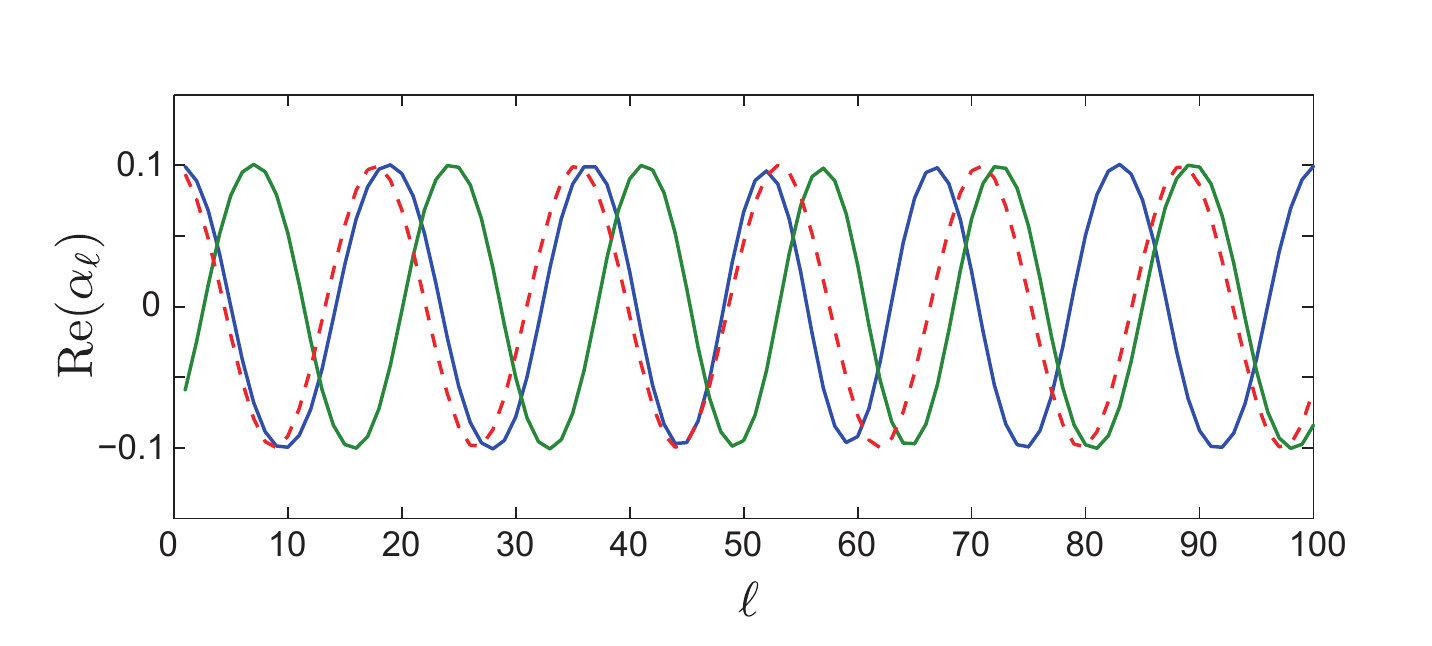}
\caption{Modulated superradiant phase. The plot shows the real part of the normalized mean field amplitude, $\alpha_\ell=\langle a_\ell \rangle/\sqrt{\mathcal{N}}$, obtained from a numerical integration of the semi-classical equations of motion~\eqref{meanfield1}-\eqref{meanfield3} for a lattice of $N_L=100$ sites with periodic boundary conditions. For this plot we have assumed $\kappa/\Delta_c=0.4$, $t/\Delta_c=0.32$ and $G/\Delta_c=0.45$, which is just above the critical coupling, $G_{\rm crit}=0.447\Delta_c$. The two solid lines show the results obtained under the same conditions, but with different random initial conditions. The dashed line shows the value of $\alpha_\ell = \cos( k_c\ell)$ with the analytically calculated critical wavevector $k_c/\pi=0.113$.}
\label{fig:Supp_Modulated}
\end{center}
\end{figure}

Both the observations that the instability of the ground state occurs at $k\neq 0$ and that the homogeneous superradiant phase is unstable above the transition in some cases discussed above indicates that for finite $\kappa$ there is a finite range of tunneling parameters $t$, where the homogeneous ansatz for $\langle a_\ell\rangle$ is invalid. Fig.~\ref{fig:Supp_Modulated} shows the stationary values of $\langle a_\ell \rangle$ in this regime for a coupling $G$, which is slightly larger than the critical coupling $G_{\rm crit}$ and for random initial conditions. We find that the values of $\langle a_\ell \rangle$ exhibit to a good approximation oscillating solutions of the form $\langle a_\ell\rangle\simeq \alpha \cos(\phi_0+k_c\ell)$, with a spontaneously chosen random offset $\phi_0$ and a wavevector $k_c$ identified above.

\subsection{C. Unstable regime}
\begin{figure}[b]
\begin{center}
\includegraphics[width=0.65\textwidth]{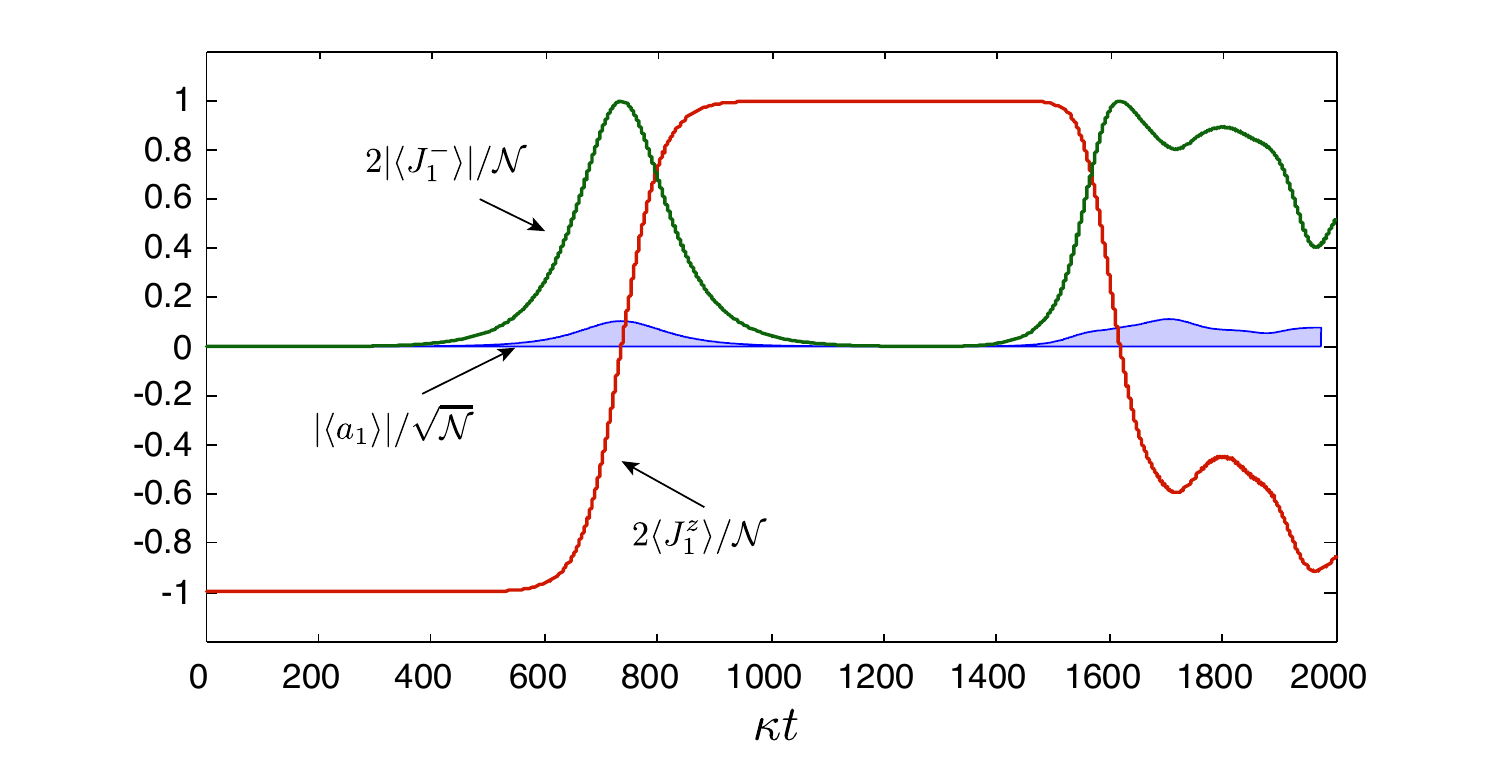}
\caption{Unstable regime. The plot shows the temporal evolution of the cavity field amplitude and the expectation values of the collective spin operators for site $\ell=1$ in a lattice of $N_L=10$ sites. For the field amplitude, only the slowly varying envelop function is shown. Additional fast oscillations are not resolved in this plot. For this simulation the parameters $\Delta_c=\Delta_s>0$, $\kappa/\Delta_c=0.4$, $t/\Delta_c=0.7$ and $G/\Delta_c=0.1$ have been used.}
\label{fig:Supp_Unstable}
\end{center}
\end{figure}

For $2t>\Delta_c$ one or more of the photonic frequencies are negative and in our model, where the spin decay rate is much smaller than all the other frequency scales, the normal phase becomes unstable for arbitrarily small values of $G$. To see this more explicitly, we change into a rotating frame with respect to $H_0=\sum_k \omega_k a_k^\dag a_k+\Delta_s \sum_ k b_k^\dag b_k$, where
\begin{equation} 
H_{\rm DLM}(t)=G\sum_k\left(b^\dagger_k a_k e^{i(\Delta_s-\omega_k)t} + b_k^\dagger a^\dagger_{-k} e^{i(\Delta_s+\omega_k)t} +{\rm H.c.}\right).
\end{equation} 
Using arguments from time-dependent perturbation theory, one sees that whenever the signs of  $\omega_k$ and  $\Delta_s$ are different, the process 
$b_k^\dag a_{-k}^\dag$ dominates over the $b_k a_{k}^\dag$ term and the collective spin modes gets excited for arbitrary small $G$.

Due to the non-linear and multi-mode character of the problem the system dynamics in the unstable regime is in general quite complex. Fig.~\ref{fig:Supp_Unstable} shows an example trajectory for the field and spin expectiation values in this regime. While for small $G$ the field in general remains small, the collective spin displays complex rotations with no significant damping over the timescales of interest.

\end{document}